\documentclass[%
 reprint,
 amsmath,
 amssymb,
 aps,
prb,
floatfix,
longbibliography
]{revtex4-1}

\usepackage[pdfencoding=auto]{hyperref}
\usepackage{graphicx}
\usepackage{color}
\usepackage[dvipsnames]{xcolor}
\usepackage{ulem}
\usepackage{subcaption}
\usepackage[utf8]{inputenc}
\usepackage{graphicx}
\usepackage[export]{adjustbox}
\usepackage{braket}
\usepackage{amsmath}
\usepackage{amssymb}
\usepackage{physics}
\usepackage{mathtools}

\usepackage{accents}
\newcommand{\dbtilde}[1]{\accentset{\approx}{#1}}

\allowdisplaybreaks

\newcommand{\Boltzmann}[1]{$\vcenter{\hbox{\includegraphics[scale = 0.3]{triangles/#1.png}}}$}

\newcommand{\leftTurn}{$\vcenter{\hbox{\includegraphics[scale = 0.3]{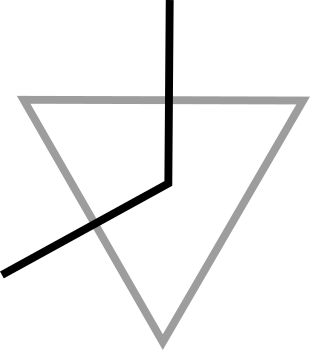}}}$}
\newcommand{\rightTurn}{$\vcenter{\hbox{\includegraphics[scale = 0.3]{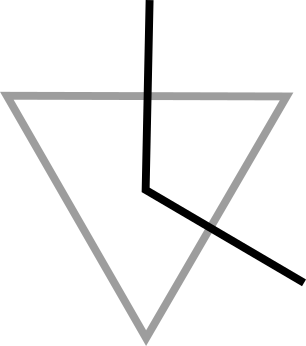}}}$}
\newcommand{\flatTurn}{$\vcenter{\hbox{\includegraphics[scale = 0.3]{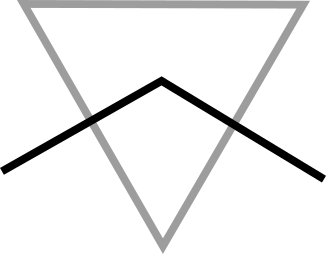}}}$}

\newcommand{\RenyiAM}[1]{\tilde{S}_{AM}^{#1}}
\newcommand{\RenyiBM}[1]{\tilde{S}_{BM}^{#1}}
\newcommand{\RenyiABM}[1]{\tilde{S}_{ABM}^{#1}}
\newcommand{\RenyiM}[1]{\tilde{S}_{M}^{#1}}

\begin{document}
\title{Statistical Mechanics of Monitored Dissipative Random Circuits}

\author{Yue (Cathy) Li}
\email{yl244@sas.upenn.edu}
\author{Martin Claassen}
\affiliation{Department of Physics and Astronomy, University of Pennsylvania, \\ Philadelphia PA 19104, USA.}%

\begin{abstract}
Dissipation is inevitable in realistic quantum circuits. We examine the effects of dissipation on a class of monitored random circuits that exhibit a measurement-induced entanglement phase transition. This transition has previously been understood as an order-to-disorder transition of an effective classical spin model. We extend this mapping to include on-site dissipation described by the dephasing and amplitude damping channel and study the corresponding 2D Ising model with generalized interactions and develop diagrammatic methods for the exact Boltzmann weights of the bonds in terms of probability of measurement $p$, the dissipation rate $\Gamma$ and the on-site Hilbert space dimension $q$. The dissipation plays the role of $\mathbb{Z}_2$-symmetry-breaking interactions, while small measurement rates reduces the ratio of the symmetry-breaking interactions to the pairwise interactions, conducive to long-range order. We analyze the dynamical regimes of the R\'{e}nyi mutual information and find that the joint action of monitored measurements and dissipation yields short time, intermediate time and steady state behavior that can be understood in terms of crossovers between different classical domain wall configurations. The presented analysis applies to monitored open or Lindbladian quantum systems and provides a tool to understand entanglement dynamics in realistic dissipative settings and achievable system sizes. \end{abstract}

\date{\today}
\maketitle

\section{Introduction}
\label{sec: intro}
A defining feature of many-body quantum systems in contrast with the classical world is the phenomenon of entanglement. The study of entanglement ranges from the cosmic-scale objects such as black holes \cite{Hayden_et_al_2016_JHEP, Swingle_2012_PRD, Happy_2015_JHEP, Xi_Harlow_Wall_PRL} to subatomic particles that constitute everyday materials \cite{Peres_1985_logic_computers, Knill_2005_QC, Lassen_2010_optical_coherence, Aolita_2015_open_system}. The entanglement entropy quantifies quantum correlations of a pure quantum state, so a detailed understanding of its behavior in realistic setups is essential to prepare states and apparatus suitable for information storage, teleportation and quantum computation. Holographic entanglement entropy has also received much attention in high energy theory, partly because it is challenging to define entanglement in a UV complete field theory \cite{Casini_2014_PRD} and partly because within AdS/CFT correspondence, the entanglement entropy of a state in a strongly coupled gauge theory is conjectured to be equal to the area of an extremal surface in its dual gravity theory \cite{RT_2006_PRL, Hubeny_2007, FLM_2013_JHEP, Happy_2015_JHEP}. 

\begin{figure}
\centering
    \subfloat[]
    {\includegraphics[width = 6cm]{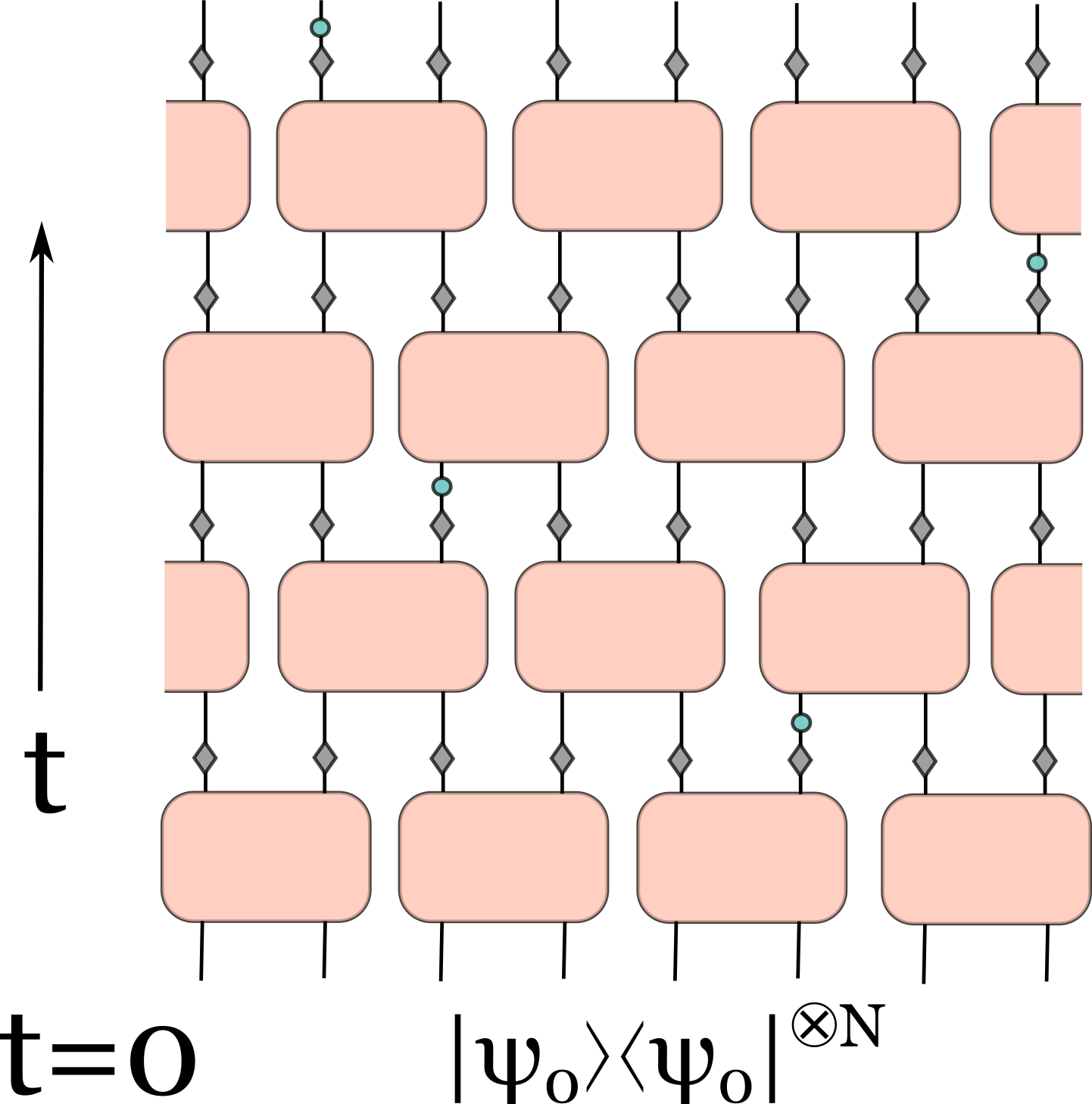}    \label{fig: brickwall}}\\  \vspace{0.5cm}
    \subfloat[]
    {\includegraphics[width = 5cm]{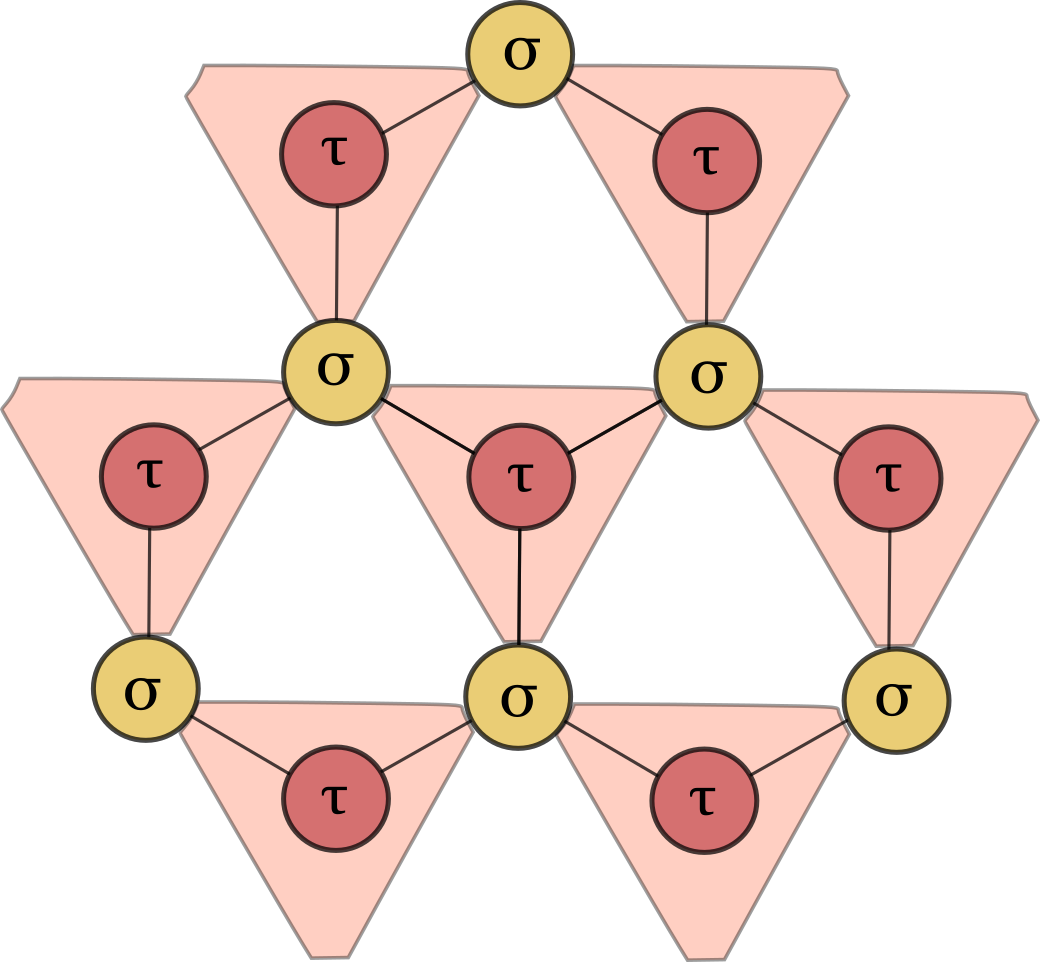}    \label{fig: star-triangle}}
    \caption{(a) Brick-wall structure of the random circuits. Each orange rectangle is a 2 qubit random Haar unitary gate. Grey diamonds denote independent dissipation with fixed rate $\Gamma$ and green circles are independent measurement with probability $p$. Ancilla are not shown in this diagram. This structure describes the system dynamics $\rho(t) = \cdots U'P \sum_i K_i U \rho(t_0) U^\dagger K_i^\dagger P U'^\dagger \cdots$. (b) The underlying lattice of the classical model that the quantum model is mapped to. The triangular lattice can be obtained using the star-triangle relationship. The time dimension becomes the vertical spatial dimension.}
\end{figure}

Recently, random quantum circuits  (FIG. \ref{fig: brickwall}) have been studied extensively due to their tractable representation of a wide range of entanglement phenomena as well as equivalences to other statistical physics problems \cite{Li_Xiao_Fisher_2019_PRB, Li_Fisher_2021_stat_qec_PRB, Nahum_Haah_Vijay_2018_PRX}. 
Random unitary circuits and driven systems can exhibit extensive entanglement entropy and ballistic information spreading \cite{Nahum_Ruhman_Vijay_Haah_2017_PRX, Nahum_Zhou_2019_PRB, Jonay_Huse_Nahum_1809_dynamics}. The resulting highly entangled state was shown to persist per quantum trajectory even when local monitored measurements are made with a sufficiently small rate, and information remains encoded non-locally. Remarkably, however, a transition from volume law to area law entanglement has been observed upon further increasing the probability of measurement over a critical threshold. This transition is known as the entanglement phase transition or measurement-induced phase transition (MPT), and has been understood in terms of a bond percolation transition through first mapping the $n$-th R\'{e}nyi entropy to the partition function of an $n!$-state Potts model on a triangular lattice (FIG.\ref{fig: star-triangle}) in the $q\to \infty$ limit, and then taking the replica limit $n\to 1$, where a Potts model turns into a bond percolation problem \cite{Lubensky_1984_Les_Houches}. In the percolation picture, the quantum volume law phase corresponds to the existence of a spanning cluster of the underlying lattice and the area law phase to the lack thereof \cite{Li_Fisher_2021_stat_qec_PRB, Nahum_Haah_Vijay_2018_PRX, Jian_You_Vasseur_Ludwig_2020_PRB}.

Realistic implementations of monitored circuits in noisy intermediate-scale quantum (NISQ) devices necessarily involve unwanted interactions with the environment. These can be equivalently regarded as subjecting the qubits to non-unitary quantum channels or Lindbladians, or as monitoring only a subset of measurements while averaging over the rest of the measurement outcomes. The volume-to-area-law transition as a function of monitored measurements disappears in the presence of dissipation \cite{Botzung_et_al_2021_PRB}, which can be understood analytically via dissipation explicitly breaking the replica symmetry that needs to be spontaneously broken in the effective classical theory \cite{Yimu_et_al_2020_PRB, Weinstein_2022_PRL, Li_Sang_Hsieh_2022, Nahum_Haah_Vijay_2018_PRX, Swingle_2021_PRL}. At the same time, an understanding of the transient entanglement dynamics prior to reaching the long-time steady state remains important, to reveal limitations for time scales and circuit depths for which useful computations can be performed. The coarse grained dynamics of random unitary circuit has been under theoretical scrutiny through the Kardar-Parisi-Zhang equation that describes the surface growth in the effective classical theory \cite{Jonay_Huse_Nahum_1809_dynamics, Nahum_Ruhman_Vijay_Haah_2017_PRX, Nahum_Zhou_2019_PRB}, and how boundary dissipation and bulk probablistic dissipation affect this dynamics is only recently studied in Clifford circuits \cite{Weinstein_2022_PRL, Liu_Li_Zhang_Jian_Yao_arxiv}.

Our work aims to tackle the effect of dissipation in a controlled manner by generalizing the classical statistical mapping of unitary circuits with measurements to  include noisy quantum channels. We compute the annealed second R\'{e}nyi entropy and mutual information by mapping the quantum circuit onto a generalized classical Ising model which includes both biasing magnetic fields and three-spin interactions, and illustrate their emergence for both dephasing and amplitude damping quantum channels. With the circuit initialized in a product state, we then study the mutual information dynamics the effective classical model at finite time with dissipation-induced symmetry-breaking contributions and analyze the dynamical regimes that emerge from domain wall configurations. We then compare the classical behavior with quantum simulations for small systems of qubits subjected to the joint action of recorded measurements and dissipation to an environment.

The rest of the paper is organized as follows: In section~\ref{sec: setup}, we introduce the setup of the ancilla-assisted dissipative plus measurement circuit. This overviews a few different representations of quantum operations, where the Lindbladian describes continuous dissipation in realistic devices, the ancilla assisted picture is instrumental in our analytical mapping, and the operator-sum representation is implemented numerically to simulate the open circuit. In section~\ref{sec: theory}, we map the second R\'{e}nyi entropy of two open circuits - dephasing and amplitude damping - to the partition function of a generalized Ising model. We study the relation between the quantum parameters and classical parameters, the former of which consists of the local Hilbert space dimension $q$, the probability of measurement $p$, the dissipation rate $\Gamma$, and the latter of which includes the external field $h$, diagonal pair-wise bonds $J_{13}$ and $J_{23}$, the horizontal pair-wise bond $J_{12}$ and the three-body interaction $J_{123}$. With the classical picture in mind, we obtain the energy-minimizing domain wall configurations from which two dynamical scales of entanglement arise for a finite system, the ballistic growth in a short time, and the exponential decay that follows. We propose a modified classical picture for the dissipative system in the thermodynamic limit, where the time scales do not depend on the system size. In section~\ref{sec: numerics}, we simulate the hybrid circuit with on-site amplitude damping and present the short-time dynamics and the steady state values of the mutual information, which confirms qualitatively the intuition and prediction of our classical theory on the 2nd R\'{e}nyi mutual information of the quantum system. Finally, we comment on possible future directions in section~\ref{sec: conclusion}.

\section{Setup}
\label{sec: setup}
The model consists of a chain of $N$ qudits initialized in a product state $\rho(t_0) = \ketbra{\psi_0}{\psi_0}^{\otimes N} = \frac 1 N (\sum_{i,j = 0}^{q-1} \ketbra{i}{j} )^{\otimes N}$, which is subjected to 2-qudit Haar random unitary gates $U$ that entangle every other pair of neighboring spins. A 2-qubit Haar operator is a matrix drawn uniformly over the unitary group $U(q^2)$.

After being subjected to a random unitary gate, each qudit interacts with its environment independently at each site. The operator-sum representation of this step is defined via the Kraus operators $\{K_i\}$: 
\begin{equation}
    \rho(t_0 + \Delta t) = \sum_i K_i \rho(t_0) K_i^\dagger.
    \label{eq: Kraus rep}
\end{equation}
The Kraus operators considered in this work can also be described by the Lindbladian evolution by integrating the following equation for finite time $\Delta t$:
\begin{equation}
\dot{\rho} = \gamma \left( \sum_i -\frac 1 2 \{ L_i L_i^\dagger, \rho \} + L_i \rho L_i^\dagger \right) \label{eq: Lindblad eq}
\end{equation}
where $\gamma$ is the dissipation rate, and $L_i$ are the jump operators. The Lindbladian assumes Markovian dissipation, with a vanishing correlation time for the bath. These two equivalent descriptions \eqref{eq: Kraus rep} and \eqref{eq: Lindblad eq} of completely positive trace-preserving channels \cite{Breuer_OpenSystem} can be transformed into each other through the Choi-Jamikolski isomorphism \cite{Simon_Felix_Kavan_2017}. We henceforth parameterize the dissipation rate via a dimensionless parameter $\Gamma$ for the Kraus operators, which can be equivalently expressed as a function of $\gamma \Delta t$ for a Lindbladian evolving the quantum state over a discrete time step $\Delta t$.

In this work, we consider two dissipation mechanisms: dephasing and amplitude damping. Dephasing can be viewed as an infinite-temperature bath that acts on the system qudits, with Kraus operators 
\begin{align*}
    K_0 &= \cos(\theta) \mathbb{I} ,\\
    K_i &= \sin(\theta) \sqrt{\frac{q}{2(q-1)}} \gamma_i,~~ i = 1,\dots,q-1.
\end{align*}
where $\gamma_i$ are the diagonal generalized Gell-Mann matrices for qudits $q$, and $[q \cos^2(\theta) - 1] / 2 = 1 - \Gamma q$ determines $\theta$.
For qubits, $q=2$, and the corresponding Lindbladian jump operator is the Pauli matrix $\sigma_z$, with $\Gamma = (1 - e^{-2 \gamma \Delta t}) / 2$. As the jump operator is Hermitian, the density matrix of the system is subjected to sequences of random unitary evolution and the dephasing channel will evolve to a fully mixed state. The derivation of the Kraus operators is included in Appendix~\ref{sec: Kraus der}.

The amplitude damping channel is described by a set of Kraus operators
\begin{align*}
    K_0 &= \ketbra{0}{0} + \sqrt{1-\Gamma} \sum_{i=1}^{q-1} \ketbra{i}{i} \\
    K_i &= \sqrt{\Gamma} \ketbra{i-1}{i},~~ i = 1,\dots,q-1.
\end{align*}
In the amplitude damping channel, the bath can be understood as a zero-temperature environment that absorbs the energy carried by photons emitted from the qudits relaxing to state $\left| 0 \right>$, or equally as a fully polarized bath of qudits. The corresponding Lindblad jump operator for $q=2$ is $\sigma_- = \frac 1 2 (\sigma_x-i \sigma_y)$. In contrast to the dephasing channel, the steady state of the Lindbladian is pure and fully polarized. Therefore, it reaches a non-trivial mixed state under joint unitary evolution and dissipation
even in the absence of measurements.

Yet another description of the open system is to introduce an ancilla that is always initialized at some fixed state $\ket{0}$ and interacts with the principal system through an interaction Hamiltonian that reproduces the dissipative dynamics after tracing out the ancilla
\begin{equation}
    \rho(t_0+\Delta t) = \Tr_{E} \left( U_\alpha \rho(t_0) \otimes  \ketbra{0}{0} U_{\alpha}^\dagger \right),
\end{equation}
where $U(t) \equiv U_{\alpha} = e^{iH_\alpha}$ describes the interaction with the environment. For example, the unitary evolution for the amplitude damping channel is generated by $H_\alpha = \alpha (\sigma_+^S \sigma_-^E + \sigma_-^S \sigma_+^E)$, where $\sigma_\pm$ are the raising/lowering operators for qudits. One can then see that the dissipation rate is $\Gamma = \sin^2{\alpha}$. This will be the primary example of dissipation considered in this manuscript.

After dissipation, the measurement of each spin in the z-direction is performed with probability $p$ independently, and the probability of an outcome is assigned according to the Born rule, $p_{q} = \frac{P_{m,q} \rho P_{m,q}}{\Tr{P_{m,q} \rho P_{m,q}}}$, where $P_{m,q}$ are projection operator on to the $q$ orthogonal states on the site $m$. The unitary-dissipation-measurement process completes one discrete time step. The next layer consists of the same procedure with Haar random unitaries shifted by one site to create the brickwork pattern in Fig. \ref{fig: brickwall}. Iterating this procedure creates an ensemble of mixed steady states with different histories of the quenched unitary $U$, the location of measurements, and measurement outcomes, distributed according to the unitary group, the probability $p$, and the Born rule respectively.

\section{Theory of the transition}
\label{sec: theory}
The MPT has been understood through a mapping between the entropic measures evaluated for $n$-replicated $(1+1)$ dimensional qudit systems to the partition function $Z$ of the $2$-d classical spin model on a triangular lattice with fixed boundary conditions \cite{Yimu_et_al_2020_PRB}. Since dissipation is equivalent to measuring and discarding the outcome, introducing dissipation does not change the essence of the mapping, so we will review the mapping here, and extend it to include dissipation. 

The primary quantities simulated in prior work are the von Neumann entropy $S(\rho_A)$ of a subset $A$ of the chain as well as the mutual information $I$ for a bipartition. The von Neumann entropy can be calculated from R\'{e}nyi entropies of order $n$: $S^{(n)}(\rho_A) = \frac{1}{1-n} \log \Tr \rho_A^n$ in the limit $n \to 1$. This quantity can be related to the partition function $Z^{(n)}$ of the $n!$-state Potts model in the $q \to \infty$ limit, so in the $n\to 1$ limit, the entanglement entropy can be understood in terms of bond-percolation or the Fortuin-Kasteleyn random cluster model \cite{Lubensky_1984_Les_Houches}. In random unitary circuits without the $q \to \infty$ and $n\to 1$ limit, a $S_n \times S_n \rtimes \mathbb{Z}^{\mathbb{H}}_2$ symmetry emerges from the replica theory and is spontaneously broken \cite{Yimu_et_al_2021_annals}. The classical spin value determines the different ways of connecting the $n$ replicas at every site, and a microstate in the spin system constitutes one complete history of the replicated qudit chain. 

The replica limit $n \to 1$ is crucial in producing the correct universality class of the measurement induced phase transition. Nevertheless, the free energy of the Ising model can still illuminate the new type of entanglement dynamics present in a dissipative monitored circuit, analogous to analyses without the replica limit in\cite{Weinstein_2022_PRL, Yimu_et_al_2021_annals}. To capture entanglement, we compute the annealed R\'{e}nyi mutual information. Although it is a classical information measure, it reveals quantum correlations as the total thermal entropy is subtracted. Beyond mutual information, deciding the separability of a mixed state is an NP-hard problem \cite{gurvits_2003_mixedNP}. Common measures for mixed state entanglement are the negativity or the logarithmic negativity\cite{Vidal_Werner_2002}. However, computing the negativity of the state requires more than two classical states per site, i.e. more than two replicas. This can be done in principle and will be simplified in $q\to \infty$ limit, see Appendix~\ref{sec: q infinity}. Empirically, the dynamics and the steady state values of mutual information and negativity are almost identical [see Appendix C], which also validates entanglement dynamics predicted from the analytic calculation of the mutual information.

In the rest of the section, we will first review how to compute the R\'{e}nyi entropy of the random circuit of qudits in which the dissipation is in the form of the dephasing channel, by mapping it to an effective spin model. The only difference between the measurement and the dephasing channel here is that after the ancilla and the on-site qudits are coupled completely, the ancilla is projected when a measurement is performed and traced out while the system dissipates. Then we compute the second R\'{e}nyi entropy of the amplitude damping channel using the Kraus operator representation. Compared to hybrid unitary circuits, the circuit with dissipation can be mapped to an Ising model with general three-body interactions on downward pointing triangles of the resulting triangular lattice, beyond the purview of exact solutions \cite{Eggarter_1975_PRB}. This mapping is exact when $n=2$, but note that for $n \geq 3$ positivity of the classical Boltzmann weights cannot be guaranteed. We then directly use the Boltzmann weights to compute possible domain wall configurations in different time regimes of the quantum system, which translates to the vertical extent of the corresponding 2D classical magnet. We finally estimate the dynamical time scales of the second R\'{e}nyi mutual information for transitions between different domain wall configurations in finite system size $N$.

\subsection{The dephasing channel}
The open quantum circuit contains a qudit chain with $q$ states $\{ \ket{1}, \ket{2}, \cdots, \ket{q} \}$ on each site and 2-qudit Haar random unitary matrices $U \in U(q^2)$ that act on neighboring spins as the orange boxes in Fig. \ref{fig: brickwall}.
The average of the von Neumann entropy \cite{nielsen_chuang_2010} over the Haar measure is simply
\begin{align}
    \mathbb{E}_U[ S_{A} ] &= -\int_U dU \Tr{ \rho_{A}(U) \log {\rho_{A}(U)}}  \nonumber \\
     & \hspace{-1cm}= \frac{\partial}{\partial n}\Big|_{ n \to 1^+} \log \mathbb{E}_U [\Tr(\rho^n)].
     \label{eq: replica}
\end{align}
The right hand side is not a properly quenched averaged $n$-th R\'{e}nyi entropy over the Haar ensemble but can be viewed as an annealed R\'{e}nyi entropy. In this paper, the ``R\'{e}nyi index" is conflated with the number of replicas, i.e. the $n \to 1$ limit will indeed give the von Neumann entropy, but without taking the limit, it does not give a properly averaged $n$-th R\'{e}nyi entropy, but an ``annealed" version of the entropy-like quantity, and is lower bounded by the $n$-th R\'{e}nyi entropy. For simplicity, it will be referred to it as the $n$-th R\'{e}nyi entropy and it should be understood that all analytic averaging stays annealed in the replica formalism until $n\to 1$ is taken. There is another similar line of replica calculation where the R\'{e}nyi index and the number of replicas are independent~\cite{Weinstein_2022_PRL, Jian_You_Vasseur_Ludwig_2020_PRB}. One can take both limits at once to obtain the von Neumann entropy, or only the replica limit to obtain the R\'{e}nyi entropies.

The unitary evolution acting on the replicas will be a tensor product of $n$ Haar matrices $U\in U(q^2)$, i.e. $U\otimes U^\dagger \otimes \cdots \otimes U^{\dagger}$, and averaging over the Haar ensemble results \cite{Collins_Haar} in a simpler network that hosts only $n!$ degrees of freedom for each $\sigma$ and $\tau \in S_{n}$ on each site
\begin{align}
    &\int_{U^{\bigotimes n}} \bigotimes_{i=1}^n dU_i\; \vcenter{\hbox{\includegraphics[scale = 0.8]{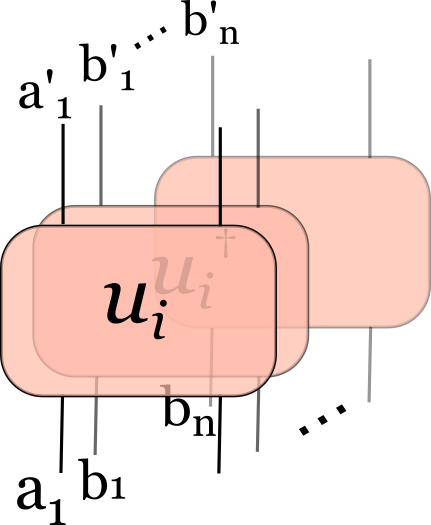}}} \nonumber \\
    &= \sum_{\sigma,\tau \in S_n} w_g(\sigma \tau^{-1},q^2 ) 
    \vcenter{\hbox{\includegraphics[scale = 0.8]{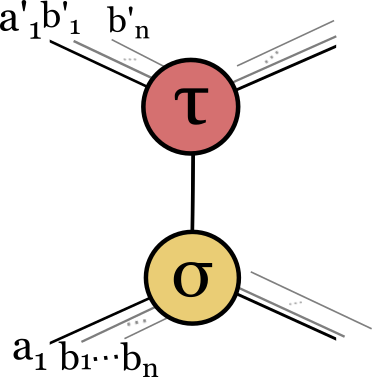}}}
    \label{eq: Haar_unitary}
\end{align}
where $w_g(\sigma \tau^{-1},q^2 ) $ is the Weingarten function, which bridges the group of unitary matrices of size $d \times d$ and the symmetric group on $q$ letters $S_q$. It is a polynomial function of $d$ of most degree $q$, defined over $S_q$ as
\begin{equation}
w_g(\sigma, d) = \frac{1}{q!^2} \sum_\lambda \frac{\chi^{\lambda}(\mathbb{I})^2 \chi^{\lambda}(\sigma)}{s_{\lambda,d}(\mathbb{I})},
    \label{eq: weingarten}
\end{equation}
where $\sigma\in S_n$, $\lambda$ is a partition, $\chi^{\lambda}$ is the character of the partition, and $s_{\lambda,d}$ is the dimension of the irreducible representation of the unitary group associated with $\lambda$~\cite{Collins_Haar}. In the quantum model, the values of $\sigma$ and $\tau$ represent different configurations of how the replica indices are connected locally. In the classical picture, they are simply the different spin states in the group $S_n$.
The unitary evolution is followed by dissipation. We choose here an ancilla representation where every qudit on the principal Hilbert space is entangled with an ancilla initialized at $\ket{0}$. The ancilla lives in a Hilbert space of $q+1$ states $\{ \ket{0}, \ket{1}, \cdots \ket{q} \}$ and becomes entangled via the unitary gate $R_{\alpha} = \sum_{i=1}^q \ketbra{i}{i} \otimes e^{-i \alpha X_i}$ where $X_i = \ketbra{i}{0} + \ketbra{0}{i}$. The strength of the interaction is controlled by the dimensionless parameter $\Gamma = \sin^2{\alpha}$ \cite{Yimu_et_al_2020_PRB}. After the application of $R_{\alpha}$, the ancilla degrees of freedom are traced out independently at every site. Tracing out the environment is averaging the state of the principal system over the ensemble of measurement outcomes, which is a simple statistical averaging and requires no replica trick.

The measurement step is operationally similar to but fundamentally different from dissipation. The ancilla gets projected onto its computational basis $\ket{i}_m$ (with each basis state describing a trajectory) which can be carried out by $\mathcal{N}_\phi[\rho] = \sum_{i=0}^q \ketbra{i}{i}_m \rho  \ketbra{i}{i}_m $. The von Neumann entropy of the subsystem $A$ can therefore be formally computed as the annealed conditional entropy between the system $M$ and $A$ \cite{Yimu_et_al_2020_PRB}: 
\begin{align}
    \mathbb{E}_U[ S_{A} ] &= \frac{\partial}{\partial n} \bigg|_{n\to 1^+} S^{(n)}_A
    \label{eq: conditionalEntropy}
\end{align}
where
\begin{align}
    S^{(n)}_A = \RenyiAM{(n)} - \RenyiM{(n)}
\end{align}
is the difference between the R\'{e}nyi entropies of the subsystems $AM$ and $M$ after the ancilla are projected onto the trajectory specified by $m$:
\begin{align}
     \tilde{S}_X^{(n)} = \log \mathbb{E}_U \Tr \{ \mathcal{N}_\phi[\rho_{X}]^n \} .
\end{align}
The average von Neumann entropy $\langle S_A \rangle$ is recovered only when $n \to 1$. However, the $S^{(n)}_A$ is substantially simpler to compute for fixed $n$. As we are interested in the understanding qualitative features of for small-system dynamics, we focus on calculating $S^{(2)}_A$ as well as a corresponding generalized mutual information measure, and compare results to small-system quantum simulations.

\subsection{Dephasing channel} \label{subsec: dephase}
Combining the above steps, the effect of dissipation followed by measurement can be compactly expressed by the Boltzmann weight of the diagonal bond of the resulting honeycomb lattice of classical effective spins, by contracting the tensor formed by the diagonal bonds with the tensor formed by the two connected spins $\sigma$ and $\tau$:
\begin{align}
    \hspace{-0.5 cm } & w^{(n)}_d(\sigma, \tau)
    = \sum_{\Vec{a},\Vec{b},\Vec{a'},\Vec{b'}} 
    \vcenter{\hbox{\includegraphics[scale = 0.7]{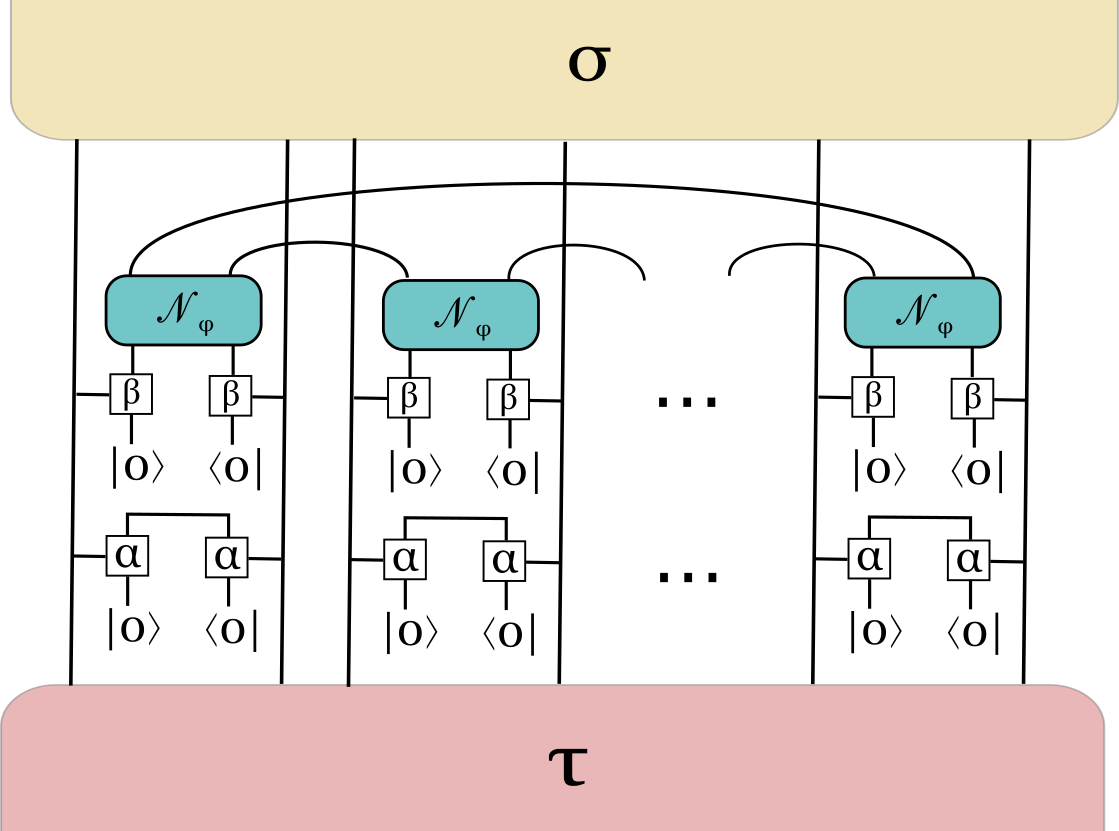}}} \nonumber\\ 
    =& \sum_{\Vec{a},\Vec{b},\Vec{a'},\Vec{b'}}M^{(n)}_{\Vec{a}\Vec{b},\Vec{a'}\Vec{b'}}    \sigma_{\Vec{a}\Vec{b}} \tau_{\Vec{a'}\Vec{b'}} \nonumber \\
    =& \sum_{\Vec{a},\Vec{b},\Vec{a'},\Vec{b'}}  \prod_{k=1}^n \Tr \Big\{\mathcal{N}_\phi \left[e^{-i \beta X_{a_k}} \ketbra{0}{0} e^{i \beta X_{b_k}^\dagger}\right] \nonumber \\
    & \hspace{0.5cm} \delta_{a_k, a'_k} \delta_{b_k, b'_k} ((1-\Gamma) + \delta_{a'_k b'_k} \Gamma) \Big\}  \sigma_{\Vec{a}\Vec{b}} \tau_{\Vec{a'}\Vec{b'}} \nonumber \\
    =& \sum_{\Vec{a},\Vec{b}} \Bigg\{ (1-p)^{n}  \prod_{k=1}^n \left((1-\Gamma) + \delta_{a'_k b'_k} \Gamma \right) \nonumber \\
    & \hspace{0.5 cm} + p^{n} \prod_{k=1}^n \delta_{a_k b_k} \delta_{b_k a_{k+1}} \left((1-\Gamma) + \delta_{a'_k b'_k} \Gamma \right) \Bigg\}  \sigma_{\Vec{a}\Vec{b}} \tau_{\Vec{a}\Vec{b}} \nonumber \\
    =& q  p^n + (1-p)^n\times  \nonumber \\
    & \hspace{-.5 cm }\sum_{\Vec{a},\Vec{b}} \sum_{k=0}^{n-1} (1-\Gamma)^k \Gamma^{n-k} q^{\mathrm{\# conn}(\prod_j \delta_{a_j b_\sigma(j)}\delta_{a_j b_\tau(j)} \prod_{\alpha}^{n-k} \delta_{a_{j_\alpha} b_{j_\alpha} } )}  \nonumber  \\
    =& q p^n + q^{\mathrm{\# cycle}(\sigma \tau^{-1})} (1-p)^n(1-\Gamma)^n  + (1-p)^n \times  \nonumber \\
    & \hspace{-.5 cm} \sum_{\Vec{a},\Vec{b}} \sum_{k=0}^{n-1} (1-\Gamma)^{k} \Gamma^{n-k} q^{\mathrm{\# conn}( \prod_{j=0}^{n-1} \delta_{a_j b_\sigma(j)}\delta_{a_j b_\tau(j)} \prod_{\alpha}^{n-k} \delta_{a_\alpha b_\alpha } )}.
    \label{eq: w_d}
\end{align} 
where $\#$conn($\cdot$) is the function that counts the number of connected components of the argument and $\#$cycle($\cdot$) counts the number of cycles of the argument. For example, fix $n=2$, $\sigma = \tau = (12)$. The exponent of $q$ in the $k=1$ term is $\# \text{conn}( \prod_{j=0}^{2-1} \delta_{a_j b_{\sigma(j)}} \delta_{a_j b_\tau(j)} \prod_{\alpha=0}^{2-1} \delta_{a_{j_\alpha} b_{j_\alpha}} =\# \text{conn}(\delta_{a_0 b_1} \delta_{a_1 b_0} \delta_{a_0 b_0} \delta_{a_1 b_1}) = 1 $. Without the $\prod_{\alpha}^{n-k} \delta_{a_\alpha b_\alpha } $ term, the connected components are the same as the number of cycles, which would be 2. The cycle structure only depends on $\sigma \tau^{-1}$ which has $S_n \times S_n$ symmetry, while the connected component depends on $\sigma$ and $\tau$ independently and there is no apparent symmetry associated with the counting of fixed points of $\sigma \tau^{-1}$. This is the replica symmetry breaking that leads to the disappearance of MPT in dissipative systems \cite{Yimu_et_al_2020_PRB, Weinstein_2022_PRL}. The tensor notation is adopted from the previous work\cite{Yimu_et_al_2020_PRB}. In the tensor network picture, the $\alpha$ gates connect all the legs between $\tau$ and $\sigma$, which makes up one connected component, and there are no more cycles. Naturally, the permutations with more fixed points, or more cycles of length 1, will have more connected components. Since the identity $\mathbb{I} \in S_n$ fixes every index, its Boltzmann weight is larger than any other spin value and the spin symmetry is broken. When $\Gamma=0$, the third term in \eqref{eq: w_d} drops out and the expression again only depends on the cycle structure of $\sigma \tau^{-1}$, restoring the symmetry. 

Because the weights from the Weingarten function can be negative, an extra step must be taken to ensure that the statistical model is sensible. Half of the spins can be summed first, resulting in a spin model defined on a triangular plaquette, such as the one in FIG.\ref{fig: star-triangle}. With dissipation, integrating out $\tau$ or $\sigma$ first can in principle, yield distinct magnetic models for the remaining spin $\sigma$ or $\tau$, respectively, while representing the same quantum behavior. We choose the former, with the Boltzmann weight of a triangular plaquette given by
\begin{align}
    w^{(n)}(\sigma_1,\sigma_2,\sigma_3) & = \sum_{\tau \in S_n} \vcenter{\hbox{\includegraphics[scale = 0.8]{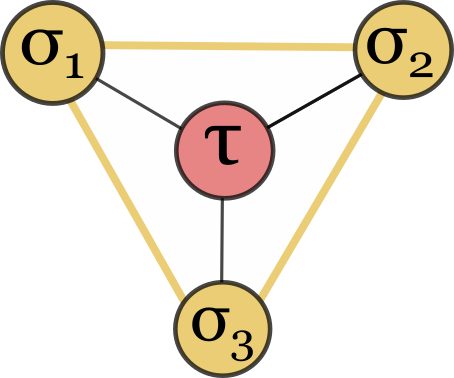}}} \nonumber\\
    & \hspace{-1cm} = \sum_{\tau \in S_n} w_g(\sigma_3 \tau^{-1},q^2 ) w_d(\sigma_1, \tau) w_d(\sigma_2, \tau).
    \label{eq: integrate_tau}
\end{align}
This is the star-triangle relation. The 3-body weight is always positive for $n=2$ and positive under certain conditions on $p$ and $q$ for general $n \geq 3$ \cite{Yimu_et_al_2020_PRB}. 

The last step that completes the mapping is to impose the correct boundary condition. The qudit chain is initialized the same way at every site and across different replicas, which means that the bottom boundary of the classical is open and its Boltzmann weight $w_{\text{bottom}}$ is 1. The boundary condition at the top is determined by the subregion $A$ for which the entropy is computed. Recall that the quantities of interest are $\RenyiAM{(n)}$ and $\RenyiM{(n)}$ after \eqref{eq: Haar_unitary} and \eqref{eq: conditionalEntropy} are combined. The former has the subregion $B$ traced out before replicating, while the latter has both $A$ and $B$ traced out before replicating. To write $S$ as a linear function of the replicated density matrix so that it can be substituted into the replica trick in \eqref{eq: replica}, one can insert $\mathbb{I}\in S_n$ in the subregion that needs to be traced out first, and insert $C^{(n)}\in S_n$ in the subregion where the R\'{e}nyi entropy needs to be computed. The first term in \eqref{eq: conditionalEntropy} can be identified with the partition function
\begin{align}
    &\mathbb{E}_U[\langle \Tr{\rho^{(n)}_{AM}} \rangle] = \mathbb{E}_U\Big[\Tr \Big\{(\mathcal{C}^{(n)}_A \otimes \mathbb{I}^{(n)}_B \otimes \mathcal{C}^{(n)}_M) \nonumber \\
    & \hspace{2 cm} \langle \rho(U)_1 \otimes \cdots \otimes \rho(U)_n \rangle \Big\} \Big]  \nonumber \\
    =& w_{\mathrm{top}}(C^{(n)}_A, \mathbb{I}^{(n)}_B, C^{(n)}_M) \sum_{\{\Vec{\sigma} \}}\prod_i w^{n}_3(\{\sigma^i_1, \sigma^i_2, \sigma^i_3\} ) \nonumber\\
    =& Z^{(n)}_{AM},
    \label{eq: partition fun}
\end{align}
where the subscripts of the partition function $Z$ means the spins in the $AM$ subregion of the boundary are pinned in a different direction from those in $B$, which are in the $\mathbb{I}\in S_{n}$ direction. For instance, when $n=2$, all the spins in the subregion $AM$ point down and the spins in the subregion $B$ point up, by convention. Therefore, combining \eqref{eq: replica}, \eqref{eq: conditionalEntropy}, \eqref{eq: partition fun}, one can compute the von Neumann entropy \eqref{eq: conditionalEntropy} as the free energy difference between two spin models with different boundary conditions:
\begin{align}
    \mathbb{E}_U [\langle S_A \rangle] &= \frac{\partial }{\partial n }\big|_{n \to 1^+ }  \left[\log Z^{(n)}_{AM}- \log Z^{(n)}_{M} \right] \nonumber \\
    &  = \lim_{n\to 1} \frac{1}{n-1}  \left[ F^{(n)}_{AM, }-F^{(n)}_{M} \right],
\end{align}
The limit is difficult to take as it requires a general expression of the right hand side as a function of $n$.  In the confined phase, which is the low-temperature limit of the classical model with the $S_n \times S_n \rtimes Z^{\mathbb{H}}_2$  symmetry, the mixed boundary condition induces at least one domain wall and the entropy hence becomes directly associated to the free energy of the domain wall. In graph theory, this domain wall is a minimal-cut surface in the graph with edges weighted by $w_g$ and $w_d$, and its existence in the tensor network construction of AdS/CFT gives a means to calculating the von Neumann entropy of the boundary subregion homologous to this surface \cite{RT_2006_PRL,Happy_2015_JHEP}.

For a mixed state, the mutual information $I = S_A + S_B - S_{AB}$ between subystems $A$ and $B$ provides a measure for quantum correlations in the system. As the limit $n \to 1$ is analytically challenging, we focus on $\tilde{S}_X^{(2)}$ as discussed above, which gives the annealed second R\'{e}nyi mutual information:
\begin{align}
    \tilde{I}^{(2)}_{A:B} &:= S^{(2)}_A + S^{(2)}_B - S^{(2)}_{A B} \nonumber \\
    & \hspace{-1.5cm} = \RenyiAM{(2)} - \RenyiM{(2)} + \RenyiBM{(2)} - \RenyiM{(2)} - \left( \RenyiABM{(2)} + \RenyiM{(2)} \right) \nonumber \\
    & \hspace{-1.5 cm} := F^{(2)}_{A} + F^{(2)}_{B} - \left( F^{(2)}_{AB} + F^{(2)}_{\emptyset} \right).
    \label{eq: mutual info}
\end{align}
The numerics has shown that the critical exponents and the critical probability of measurement do not depend on the R\'{e}nyi index \cite{Skinner_Ruhman_Nahum_2019_PRX}. The calculation, therefore, simplifies to approximating the free energies of the Ising model $F_X \equiv \tilde{S}_X^{(2)}$ as a function of the quantum parameters in different limits of the system size and time, for different boundary conditions that are determined by $X$. We take $A$ and $B$ to be bipartitions of the top boundary of the statistical model and the subregion $M$ does not need to appear since it is fixed in the direction $\mathbb{I}\in S_{n}$ in every term. Furthermore, $F_{\emptyset} =S^{(n)}_M$ represents the spin model with the entire top boundary fixed to be $\mathbb{I}\in S_{n}$. This term is necessary to compute the conditional entropy and it functions as a constant shift in the mutual information. We note that the R\'{e}nyi entropies and its generalized annealed versions do not satisfy subadditivity, so the corresponding mutual information can be negative \cite{Scalet_2021_Renyi}. Alternative generalizations of mutual information have been proposed such as the Petz R\'{e}nyi mutual information or the R\'{e}nyi divergence \cite{Scalet_2021_Renyi, Jonah_2023_PRL}, the study of which will be left for future work.

\begin{figure}[ht]
    \centering
    \includegraphics[width = 8cm]{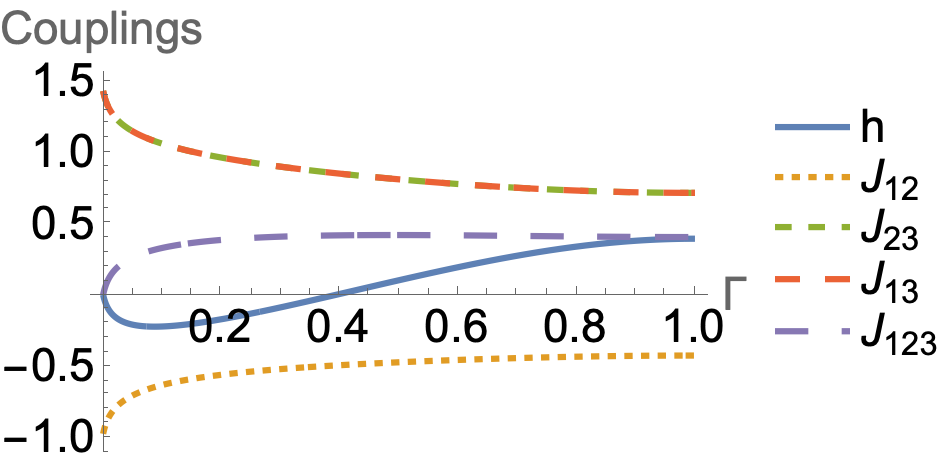}
    \caption{Classical parameters as a function of the parameters in the quantum model, with $p=0.1, q=2$.}
    \label{fig: depolarzation at p=0.1, q=2}
\end{figure}

\begin{figure}[ht]
    \centering
    \includegraphics[width = 8cm]{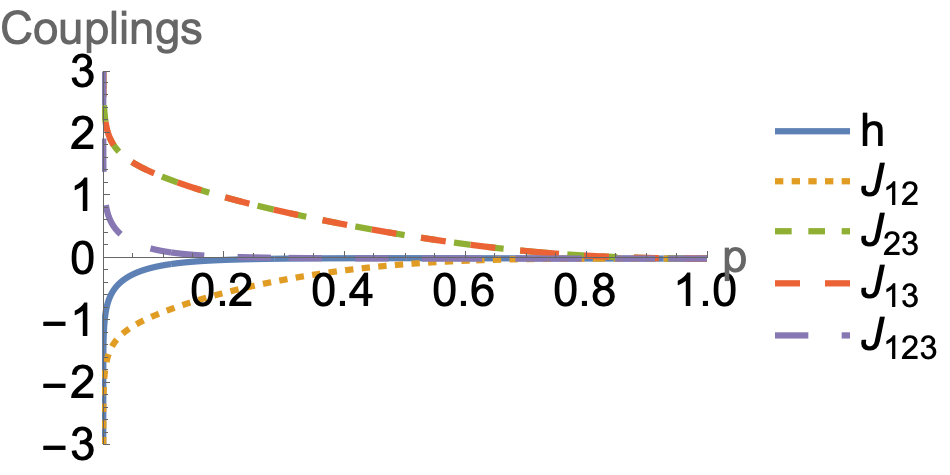}
    \caption{Classical parameters as a function of the parameters in the quantum model, with $\Gamma=0.001, q=2$.}
    \label{fig: dephasing at Gamma=0.001, q=2}
\end{figure}

With the boundary conditions established, we are now ready to map the twice-replicated open quantum circuit ($n=2$) above to the Ising spins defined the downward pointing triangular plaquette with arbitrary interactions. The Boltzmann weight on one plaquette is:
\begin{align}
    & w(\sigma_1,\sigma_2,\sigma_3) = e^{-\beta E(\sigma_1,\sigma_2,\sigma_3)} \nonumber \\
    &:= \exp\big\{ h_0+ h_1\sigma_1 + h_2\sigma_2 + h_3\sigma_3 + J_{12} \sigma_1 \sigma_2 \nonumber \\
    & \hspace{1cm} + J_{23} \sigma_2 \sigma_3 + J_{31} \sigma_3 \sigma_1 + J_{123} \sigma_1 \sigma_2 \sigma_3\big\},
    \label{eq: ising plaquette}
\end{align}
Here, $J_{123}$ is the coupling between the three spins for a downward-facing triangle, the $J_{ij}$ are the pairwise couplings, $h_i$ describe effective external magnetic fields, and $h_0$ is a normalization. Each of these effective parameters depends on the measurement and dissipation rates $p$ and $\Gamma$, as well as the qudit dimension $q$. All eight parameters are defined via the weights for eight distinct configurations of triangle spins $w(\sigma_1,\sigma_2,\sigma_3)$, where $\sigma_1,\sigma_2$, and $\sigma_3$ are taken to denote the top left, top right and bottom spin of downward-pointing triangles. Combining the triangles into a triangular-lattice Ising model, the total magnetic field acting on each spin is $h = \sum_{i=1}^3 h_i$ because every lattice spin is at the corner of three downward triangles.

We illustrate how the classical parameters behave as one adjusts the quantum parameters $\Gamma$ and $p$ respectively in FIG.\ref{fig: depolarzation at p=0.1, q=2} and FIG. \ref{fig: dephasing at Gamma=0.001, q=2}. In the absence of dissipation, all the symmetry-breaking terms $h_i$ and $J_{123}$ are 0, which recovers the spin model in \cite{Yimu_et_al_2020_PRB}. Increasing the probability of measurement $p$ weakens pairwise interactions, eventually reaching a transition from ferromagnetic order to disorder.
 When $\Gamma > 0$, the symmetry-breaking terms $h_i$ and $J_{123}$ become non-zero, and the probability of measurement $p$ decreases bias terms at a much faster rate than the pairwise interaction couplings (FIG.\ref{fig: dephasing at Gamma=0.001, q=2}). We will see similar effects of $p$ and $\Gamma$ for the amplitude damping channel in the next subsection.

\subsection{Amplitude damping channel}
\label{sec: analytical sigma- channel}

Now consider the amplitude damping channel. The reader who only wishes to know the consequence of the calculation, i.e. the entanglement dynamics described by the domain walls rather than the details of using of replica symmetry breaking tensor diagrams, can skip to section~\ref{sec: mincut}. This channel is commonly used to model spontaneous emission processes, e.g., the qudit chain is coupled to a polarized zero-temperature bath, which forces the system qudits to collapse to $\ket{0}$. The Kraus operators for the channel are $\{K_0 = \ketbra{0}{0} + \sqrt{1-\Gamma} \sum_{i=1}^{q-1} \ketbra{i}{i}, K_m = \sqrt{\Gamma} \ketbra{m-1}{m} \}_{m=1}^{q-1}$. To introduce these to the mapping, we replace the ancilla couplings ($\alpha$ gates) by a sum over Kraus operators. The weights are computed from contracting $\sum_{m=0}^{q-1}  K_{m, a'_i a_i} K^{\dagger}_{m, b'_i b_i}$, $i=1, \dots, n$ with the $\sigma$ and $\tau$ tensors, weighted by the same coupling to the measurement ancillae. One obtains the following:
\begin{widetext}
\begin{align}
    \hspace{-0.5 cm } & w^{(n)}_d(\sigma, \tau)
    = \sum_{\Vec{a},\Vec{b},\Vec{a'},\Vec{b'}} 
    \vcenter{\hbox{\includegraphics[scale = 0.8]{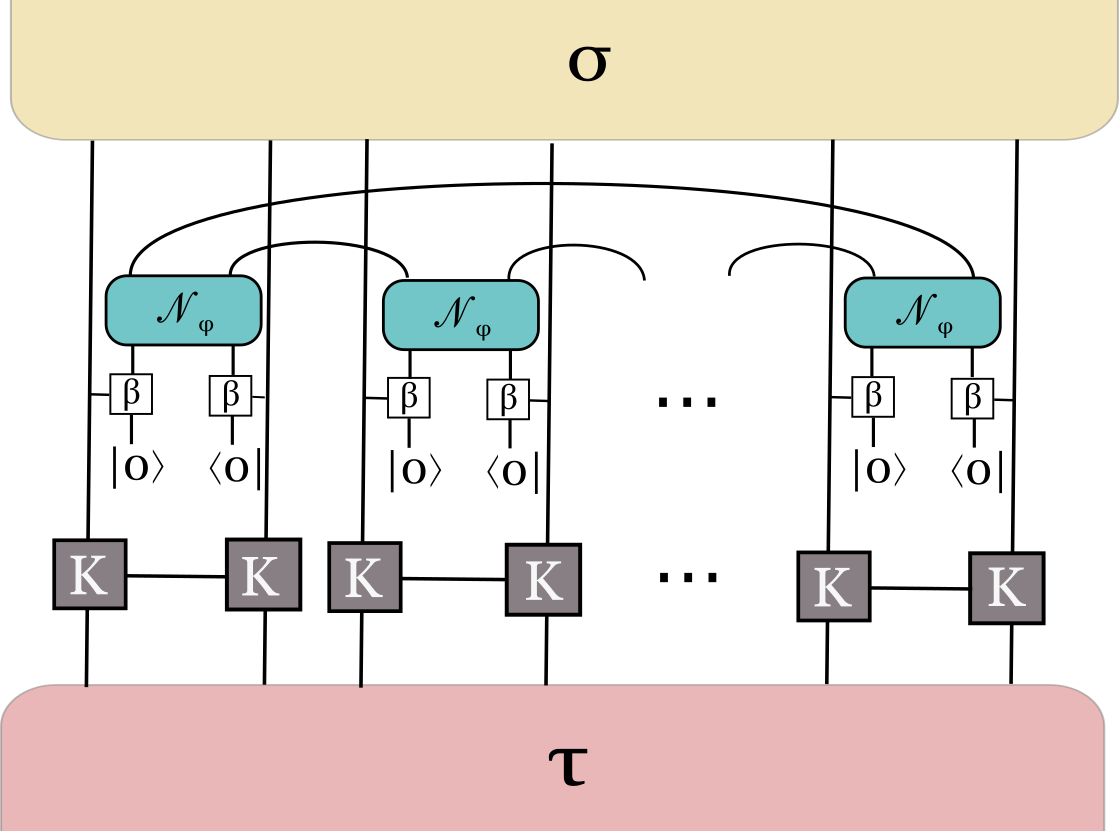}}} \nonumber \\
    &= \sum_{\Vec{a},\Vec{b},\Vec{a'},\Vec{b'}} \prod_{j=1}^n \Big[(\delta_{a'_j 0} \delta_{a_j 0} + \sqrt{1-\Gamma} (\delta_{a'_j a_j} - \delta_{a'_j 0} \delta_{a_j 0}))(\delta_{b'_j 0} \delta_{b_j 0} + \sqrt{1-\Gamma} (\delta_{b'_j b_j} - \delta_{b'_j 0} \delta_{b_j 0}) ) \nonumber \\
    & \hspace{1cm}+ \Gamma \sum_{m=1}^q \delta_{a'_j m-1} \delta_{a_j m}\delta_{b'_j m} \delta_{b_j m-1} \Big]\Big[(1-p)^n + p^n \prod_{k=1}^n \delta_{b'_k a'_{k+1}} \delta_{a'_k b'_k}\Big] \delta_{a'_j b'_{\sigma(j)}}\delta_{a_j b_{\tau(j)}} \nonumber \\
    \label{eq: w_d of sigma_-}
\end{align}
\end{widetext}

The first bracket in the product \eqref{eq: w_d of sigma_-} explicitly breaks the replica symmetry. In the dephasing channel, the identity $\mathbb{I}\in S_n$ will give the highest Boltzmann weight because it has the most connected components. In this channel, the same structure is imposed by the $K_i$ terms, favoring the configuration $\mathbb{I}$ that maximizes the number of total connections of the delta functions. To simplify the weights from \eqref{eq: w_d of sigma_-}, we create the following diagrammatic rules. The terms from $K_0$ has the two following contributions - the stub for $\delta_{a_i 0}$s and the legs for $\delta_{a_i a'_i}$s: 
\begin{align}
\delta_{a_i 0} \delta_{b_i 0} \delta_{a'_i 0} \delta_{b'_i 0} &= \vcenter{\hbox{\includegraphics[scale = 1]{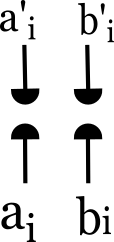}}},\;\nonumber \\
\sum_{a_i, a'_i, b_i, b'_i} \delta_{a_i 0} \delta_{b_i 0} \delta_{a_i 0} \delta_{b_i 0} &=
\vcenter{\hbox{\includegraphics[scale = 1]{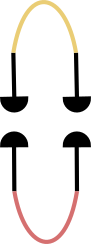}}} =1
\label{eq: K1 term1}
\end{align}
where the stub is contracted with the state $\ket{0}$, which gives the sum 1. The straight legs live in the $(q-1)$ dimensional Hilbert space with the states $\{1, \dots, q-1 \}$ 
\begin{align}
    \delta_{a_i a'_i} \delta_{b_i b'_i} = \vcenter{\hbox{\includegraphics[scale = 1]{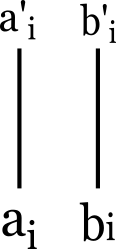}}};  \; \sum_{a_i, a'_i, b_i, b'_i=1}^{q-1} \delta_{a_i a'_i} \delta_{b_i b'_i} \vcenter{\hbox{\includegraphics[scale = 1]{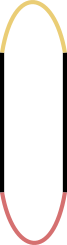}}} = q-1.
    \label{eq: K1 term2}
\end{align}
Each term above is accompanied by a factor of $(1-\Gamma)$. The contraction of \eqref{eq: K1 term1} and \eqref{eq: K1 term2} between different replicas yields 0. The tensor and tensor contraction associated with the operator $K_i$ is
\begin{align}
\delta_{a_i b_i} \delta_{a'_i b'_i} \delta_{a_i a'_i-1} &= \vcenter{\hbox{\includegraphics[scale = 1]{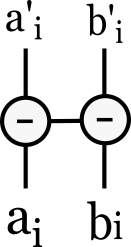}}}; \nonumber \\
\sum_{a_i, b_i =1}^{q-1} \sum_{a'_i b'_i =0}^{q-1} \delta_{a_i b_i} \delta_{a'_i b'_i} \delta_{a_i-1 a'_i} &= \vcenter{\hbox{\includegraphics[scale = 1]{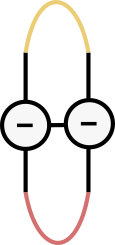}}} = q-1,
\label{eq: K2 term}
\end{align}
where the circles with the minus sign ensure that $a'_i = a_i -1 $ and $b'_i = b_i -1$. The diagram is always accompanied by $\Gamma$. The upper legs of the circle live in the Hilbert space spanned by $\{\ket{0}, \dots, \ket{q-2}\}$ and the lower legs by $\{\ket{1}, \dots, \ket{q-1}\}$, both having one fewer dimension than the straight legs from $K_0$. This means that in the cross terms, the contraction between the Hilbert space $\{\ket{0}, \dots, \ket{q-2}\}$ of $K_0$ and $\{\ket{1}, \dots, \ket{q-1}\}$ of $K_i$ gives $q-2$. 

The contractions shown above occur when the output ket $a'_i$ and bra $b'_i$ are identified within each replica, i.e. when the spin points in the $\uparrow$ direction. From this, we can calculate the diagrams that contribute to the weight $w(\uparrow \uparrow)$ in the Ising case ($n=2$)  as the following:
\begin{align}
    w_d(\uparrow \uparrow) & = \Bigg[ \vcenter{\hbox{\includegraphics[scale = 0.8]{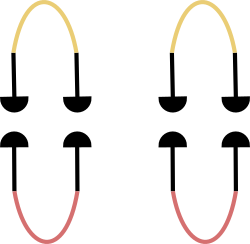}}}  + 2 (1-\Gamma) \vcenter{\hbox{\includegraphics[scale = 0.8]{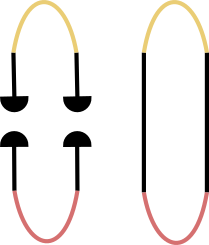}}} + (1-\Gamma)^2 \vcenter{\hbox{\includegraphics[scale = 0.8]{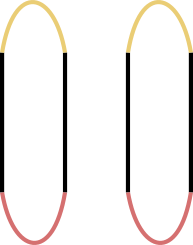}}} \nonumber \\
    & \hspace{-1cm} + 2 \Gamma\vcenter{\hbox{\includegraphics[scale = 0.8]{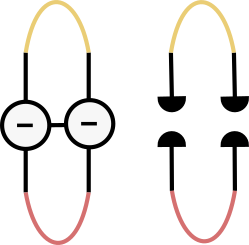}}} +  2 \Gamma (1-\Gamma) 
    \vcenter{\hbox{\includegraphics[scale = 0.8]{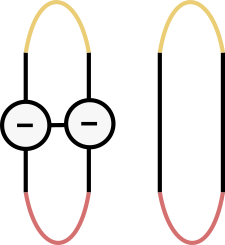}}} +  \Gamma^2 \vcenter{\hbox{\includegraphics[scale = 0.8]{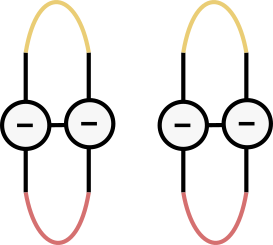}}} \Bigg] (1-p)^2 \nonumber \\
    & \hspace{-1cm}  + \Bigg[ \vcenter{\hbox{\includegraphics[scale = 0.8]{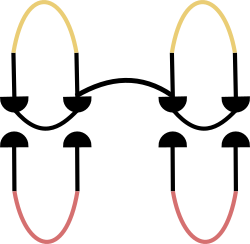}}} + (1-\Gamma)^2 \vcenter{\hbox{\includegraphics[scale = 0.8]{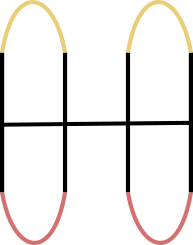}}} + 2 \Gamma \vcenter{\hbox{\includegraphics[scale = 0.8]{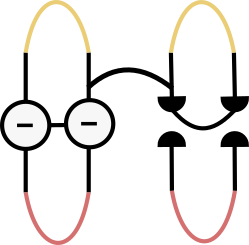}}} \nonumber \\
    & \hspace{-1cm} +  \Gamma^2\vcenter{\hbox{\includegraphics[scale = 0.8]{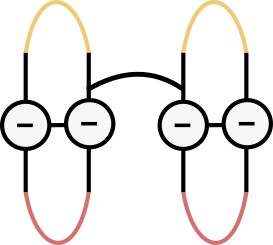}}} 
    \Bigg] p^2 \nonumber \\
    & \hspace{-1cm} = [1 + 2(1-\Gamma)(q-1 ) + (1-\Gamma)^2 (q-1)^2 + 2\Gamma(q-1 ) \nonumber \\
    & \hspace{-0.5cm} + 2\Gamma (1-\Gamma)(q-1)^2 + (q-1)^2 \Gamma^2](1-p)^2 \nonumber \\
    & \hspace{-0.5cm} + \big[1 + (1-\Gamma)^2(q-1)  + 2\Gamma  + \Gamma^2
    \big]p^2 \nonumber \\
     & \hspace{-1cm}  = q^2 (1 - 2 p +p^2) + q p^2 + 2 p^2  \Gamma^2.
\end{align}
Factors of 2 are introduced due to the invariance of the tensor contraction under the exchange of the two replicas. When the brakets connect different copies, as a general rule, any connected component contributes a factor of $q-1$, except for those involving a stub \eqref{eq: K1 term1}, which contributes 1, and when the \eqref{eq: K1 term2} is connected to the upper leg of \eqref{eq: K2 term}, which contributes $q-2$. The other weights can be derived similarly:
\begin{align}
     w_d(\downarrow \downarrow) & =  \Bigg[ \vcenter{\hbox{\includegraphics[scale = 0.8]{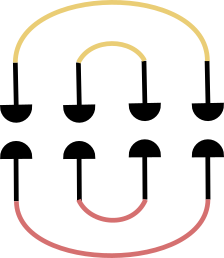}}} + (1 - \Gamma)^2 \vcenter{\hbox{\includegraphics[scale = 0.8]{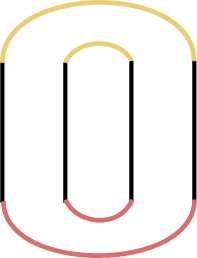}}} + 2(1 - \Gamma) \vcenter{\hbox{\includegraphics[scale = 0.8]{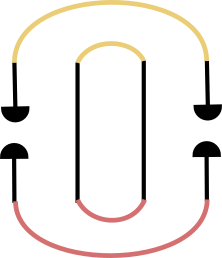}}} \nonumber \\
     & \hspace{-1cm} + \Gamma^2 \vcenter{\hbox{\includegraphics[scale = 0.8]{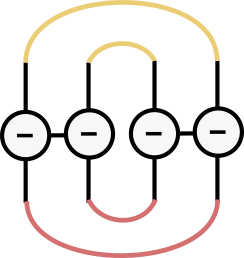}}}\Bigg]
     (1-p)^2 + \Bigg[ \vcenter{\hbox{\includegraphics[scale = 0.8]{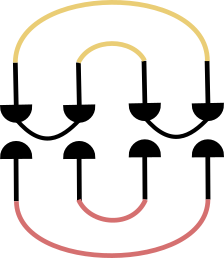}}} + (1-\Gamma)^2 \vcenter{\hbox{\includegraphics[scale = 0.8]{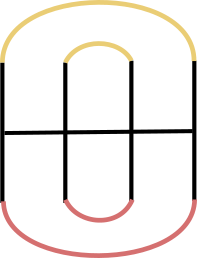}}} \nonumber\\
     & \hspace{-1cm} + \Gamma^2 \vcenter{\hbox{\includegraphics[scale = 0.8]{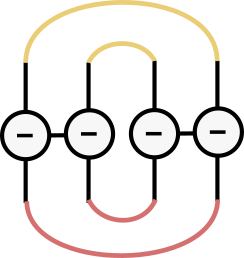}}}\Bigg] p^2 \nonumber \\
     & \hspace{-1cm} = (q^2 (1-\Gamma)^2 + q(2 - \Gamma) \Gamma) (1-p)^2 \nonumber \\
     & + (1+(q-1)(2\Gamma^2 - 2\Gamma +1))p^2 \\~\nonumber\\
    w_d(\uparrow \downarrow) & = \Bigg[ \vcenter{\hbox{\includegraphics[scale = 0.8]{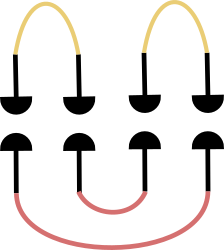}}} +(1-\Gamma)^2 \vcenter{\hbox{\includegraphics[scale = 0.8]{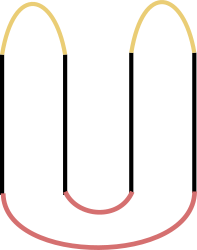}}} \nonumber \\
     & \hspace{-1cm} + 2\Gamma(1-\Gamma)\vcenter{\hbox{\includegraphics[scale = 0.8]{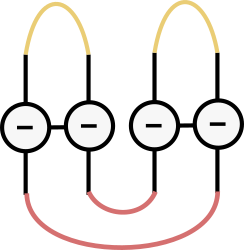}}} +\Gamma^2 \vcenter{\hbox{\includegraphics[scale = 0.8]{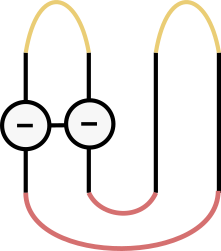}}} \Bigg] (1-p)^2 \nonumber \\
     &\hspace{-1cm} + \Bigg[ \vcenter{\hbox{\includegraphics[scale = 0.8]{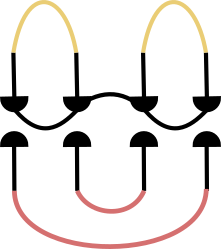}}} + \vcenter{\hbox{\includegraphics[scale = 0.8]{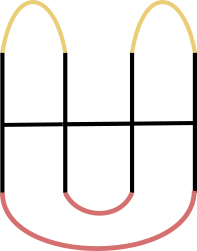}}} (1-\Gamma)^2 +\vcenter{\hbox{\includegraphics[scale = 0.8]{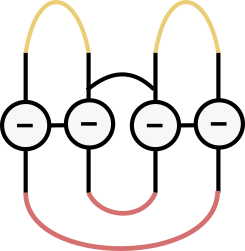}}} \Gamma^2 \Bigg] p^2 \nonumber \\
     &\hspace{-1cm} =  [1+(q-1) (1-\Gamma)^2 + (q-1) \Gamma^2 + 2(q+1)\Gamma(1-\Gamma) ]  \nonumber \\
     & (1-p)^2 +[1+(q-1) (1-\Gamma)^2 + (q-1)\Gamma^2] p^2 \nonumber \\
     &  \hspace{-1cm} = q ((1-p)^2 + p^2(2\Gamma^2- 2\Gamma +1))+2p^2 \Gamma(1-\Gamma)\\~\nonumber\\
     w_d(\downarrow\uparrow) & =  \Bigg[ \vcenter{\hbox{\includegraphics[scale = 0.8]{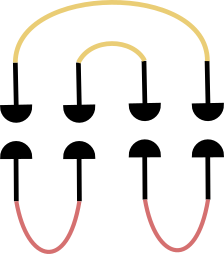}}} + (1-\Gamma)\vcenter{\hbox{\includegraphics[scale = 0.8]{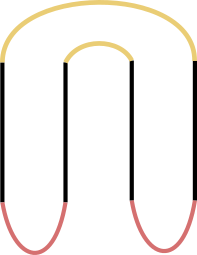}}} + \Gamma^2 \vcenter{\hbox{\includegraphics[scale = 0.8]{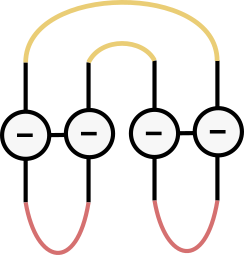}}}  \nonumber \\
     & \hspace{-1cm} + 2\Gamma (1-\Gamma) \vcenter{\hbox{\includegraphics[scale = 0.8]{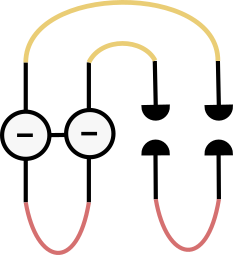}}} + 2 \Gamma \vcenter{\hbox{\includegraphics[scale = 0.8]{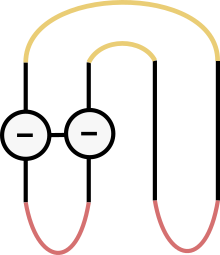}}} \Bigg] ((1- p)^2 +p^2) \nonumber \\
     &  \hspace{-1cm} = [1 + (q-1)(1-\Gamma)^2 + (q-1)\Gamma^2 \nonumber \\
     & + 2(q-2) \Gamma (1-\Gamma) ] (p^2 + (1-p)^2) \nonumber \\
     &  \hspace{-1cm} = (q + 2 \Gamma^2) (1-2p + 2p^2) 
     .
\end{align}
We note that the diagrams illustrate how to generalize the weight computation for general $n> 2$, but the positivity of the Boltzmann weights is again not guaranteed.

At small values of $\Gamma$ and $p$, the Ising model representation of the emission channel behaves similarly to the dephasing channel. This can be understood as the principal function of dissipation at small $\Gamma$ is making the quantum state mixed regardless of the specific dissipative mechanism. At small enough $\Gamma$, dissipation increases the biasing terms and decreases pair-wise interactions, as can be seen from FIG.\ref{fig: amplitude damping (h,J) in terms of Gamma}. As $\Gamma$ increases, the bath ultimately forces the quantum system into a fully polarized state. At very large $\Gamma$, the parameters of the two classical models have very different behaviors, which illustrates that the two steady states, $\frac 1 N \mathbb{I}^{\otimes N}$ and $\ket{0}^{\otimes N}$, of the open quantum circuits are drastically different, as a result from interacting with an infinite temperature and zero-temperature bath.

\begin{figure}[ht]
    \centering
    \includegraphics[width = 8cm]{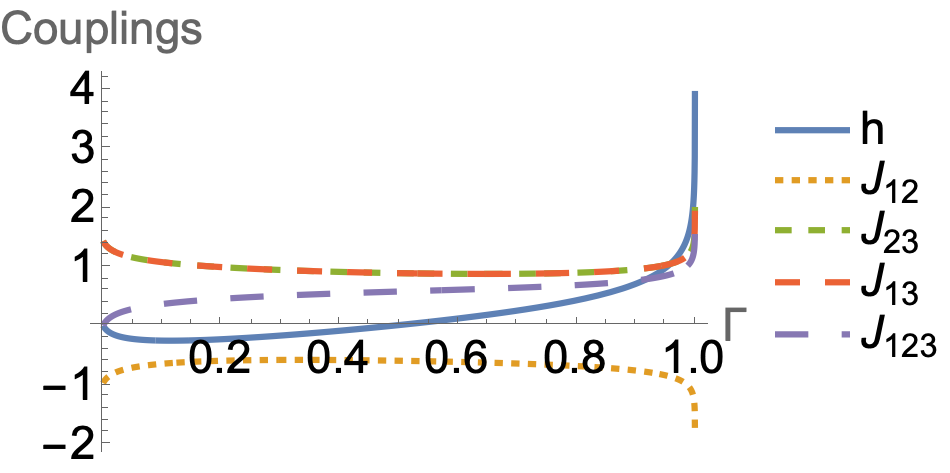}
    \caption{Classical parameters as a function of the parameters in the quantum model in the amplitude damping model, with $p=0.1, q=2$. }
    \label{fig: amplitude damping (h,J) in terms of Gamma}
\end{figure}

\begin{figure}[ht]
    \centering
    \includegraphics[width = 8cm]{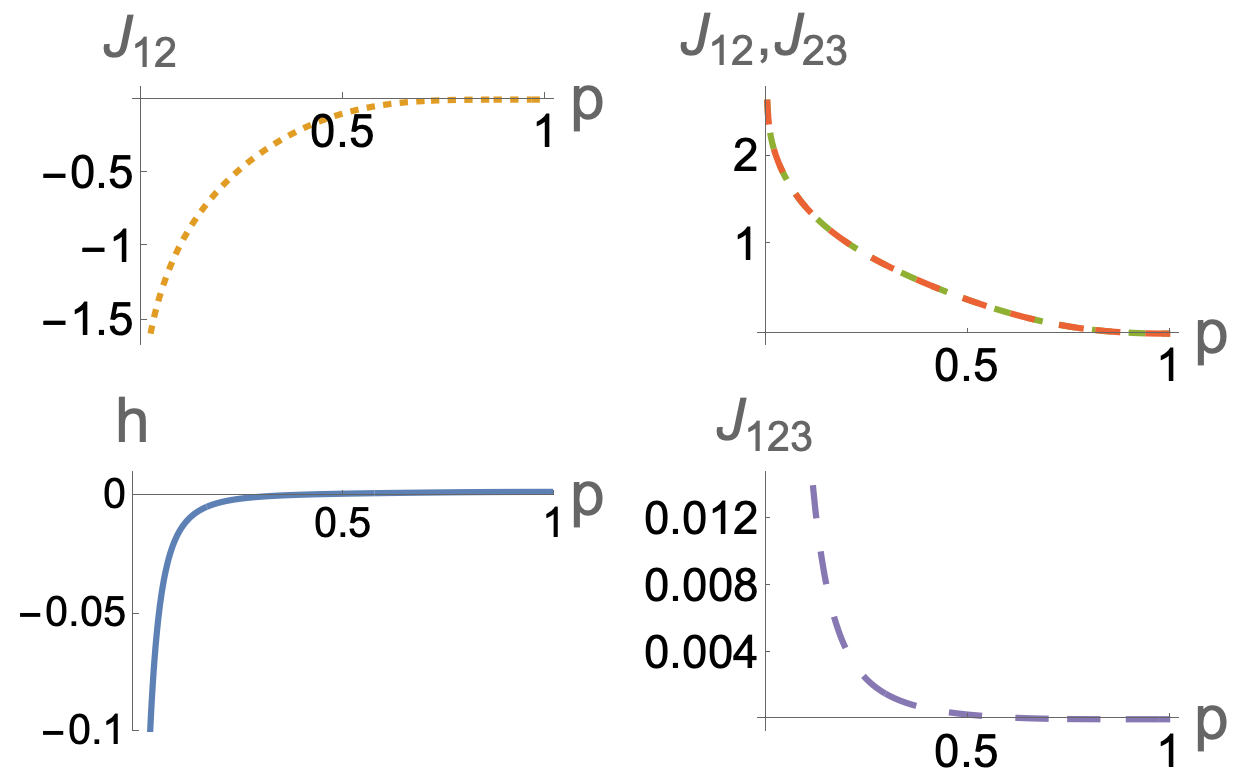}
    \caption{Classical parameters as a function of the parameters in the quantum model in the amplitude damping model, with $\Gamma=0.001, q=2$.}
    \label{fig: amplitude damping (h,J) in terms of p, Gamma=0.001, q=2}
\end{figure}

The dependence of the individual couplings on $p$ at $\Gamma =0.001$ are shown in FIG.\ref{fig: amplitude damping (h,J) in terms of p, Gamma=0.001, q=2}. As $p$ increases, the symmetry-breaking $h$ and $J_{123}$ are quickly suppressed to orders of magnitude smaller than $J_{ij}$, which restores the long-range order in the pairwise Ising model on the triangular lattice \cite{Yimu_et_al_2020_PRB, Eggarter_1975_PRB} and nudges the dissipative system towards larger steady-state entanglement. This suggests sweet spots which, although obeying area laws for mutual information and entanglement negativity, yield maximal prefactors for their asymptotic behavior by virtue of competition between dissipation and measurements. The consequences for the mutual information will be further analyzed in the next section.

\subsection{\texorpdfstring{$\tilde{S}^{(n)}_X$}{Lg} and domain walls }
\label{sec: mincut}

We now turn to extract the mutual information $\tilde{I}^{(2)}$ from domain walls that form for different boundary conditions in  \eqref{eq: mutual info}. Without dissipation, the free energy of domain walls scales with the subsystem size $N/2$ in the small-$p$ ferromagnetic phase for $F_A$ and $F_B$ while $F_{AB}$ and $F_{\emptyset}$ remain trivial, which results in an entanglement volume law.

At finite $\Gamma$, the situation becomes richer: the energy of boundary-enforced domains with spins that are anti-aligned to the bias field and three-spin interaction now scales with the domain area rather than just the domain wall length. The enclosed area is a product of the subsystem size and the time-like direction, and preferable domain wall configurations will therefore sensitively depend on time. Furthermore, $F_{AB}$ can favor hosting domain walls as well. The expected area laws in the steady state must consequently result from the cancellation of free energies from different boundary conditions.

\subsubsection{ \texorpdfstring{$p=0$}{Lg} measurement limit, \texorpdfstring{$t \to \infty$}{Lg} }
\label{sec: p=0 msm limit}

\begin{figure}
    \centering
    \includegraphics[width = 8cm]{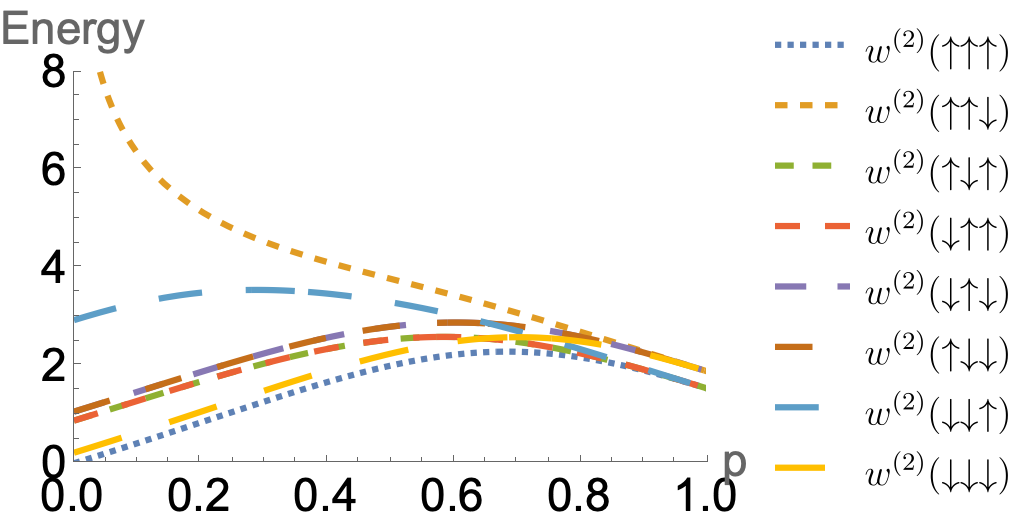}
    \caption{The energy of each plaquette in the Ising model \eqref{eq: partition fun} mapped from the dissipative open circuit. $\Gamma$ is set to 0.1 to exaggerate the Ising symmetry breaking in the weights.}
    \label{fig: Boltzmann weights}
\end{figure}

We start by reviewing the limit of an unmonitored circuit at long times and use the exact dependence of the Boltzmann weights on $(p,\Gamma)$ to deduce the saddle point of the classical model with different boundary conditions and in the two-dimensional limit for the classical model, to extract the steady state behavior of the mutual information. This is tractable as there is no entropic contribution to the free energy of the model. At $p=0, \Gamma \neq 0$, the classical system will minimize the area spanned by $\vcenter{\hbox {\includegraphics[scale = 0.3]{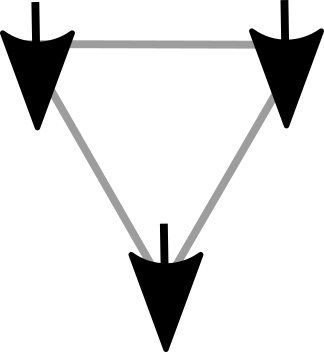}}}$, because $\vcenter{\hbox {\includegraphics[scale = 0.3]{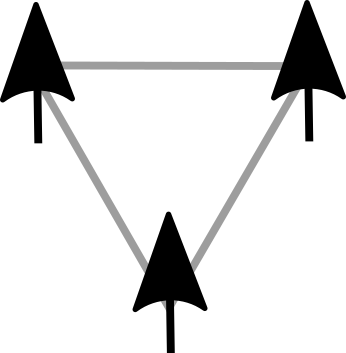}}}$ will always cost less energy from FIG.\ref{fig: Boltzmann weights}. Even though domain walls are energetically costly, in the thermodynamic and at long times, the bulk will be fully $\uparrow$ polarized regardless of the boundary condition. All domain walls therefore stick to the top boundary and determine the R\'{e}nyi mutual information \eqref{eq: mutual info}. After subtracting a system-spanning domain wall in $F_{AB}$ from two half system domain walls in $F_{A}$ and $F_{B}$, only a contribution from the boundary between subsystems $A$ and $B$ remains. This means that the mutual information obeys an area law, as expected. We carry out the infinite $q$ expansion of the domain wall configuration explicitly in Appendix.\ref{sec: q infinity}. From table. \ref{table: boltzmann}, the leading order of $\frac{1}{q}$ to the R\'{e}nyi mutual information is:
\begin{align}
    \tilde{I}^{(2)}_{A:B} &=  F^{(2)}_{A} + F^{(2)}_{B} - (F^{(2)}_{AB} + F^{(2)}_{\emptyset} ) \nonumber \\
    & \hspace{-1.5 cm} = E(\downarrow \in A, \; \uparrow\in B) + E(\uparrow \in A, \; \downarrow\in B) \nonumber \\ & \hspace{1cm} -(E(\downarrow \in A B) + E(\uparrow\in A B)) \nonumber \\
    & \hspace{-1.5 cm} = F\left(\vcenter{\hbox {\includegraphics[scale = 0.5]{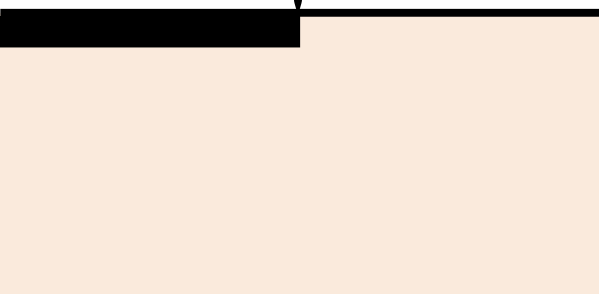}}} \right)+ F\left(\vcenter{\hbox {\includegraphics[scale = 0.5]{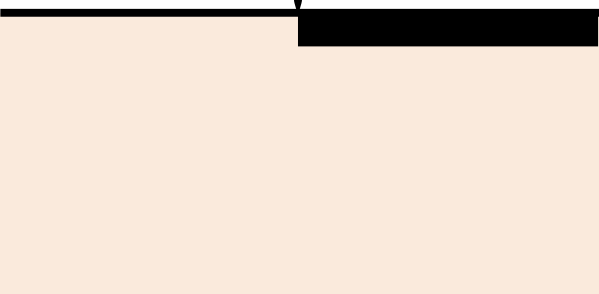}}}\right) \nonumber \\
    & - F\left(\vcenter{\hbox {\includegraphics[scale = 0.5]{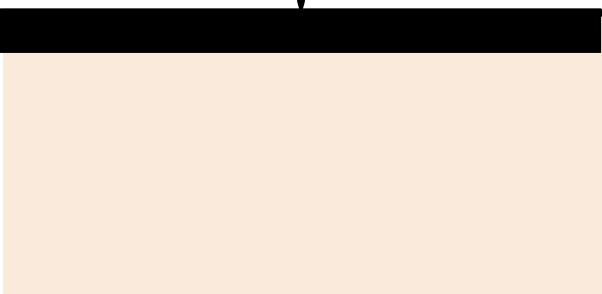}}}\right) - \left(\vcenter{\hbox {\includegraphics[scale = 0.5 ]{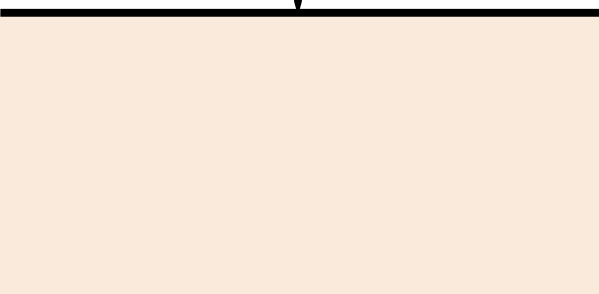}}} \right) \nonumber \\
    & \hspace{-1.5 cm} \approx 2  \lim_{q\to \infty}\bigg( E(\vcenter{\hbox {\includegraphics[scale = 0.4]{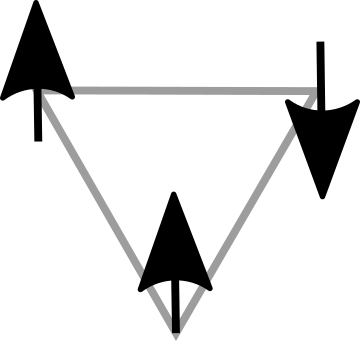}}}) + E(\vcenter{\hbox {\includegraphics[scale = 0.4]{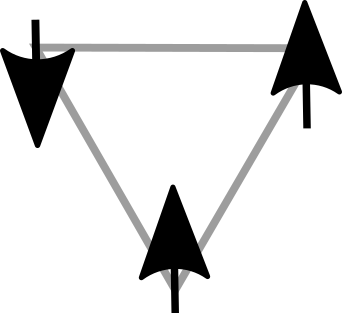}}}) \nonumber \\
    & \hspace{1 cm} - E (\vcenter{\hbox {\includegraphics[scale = 0.4]{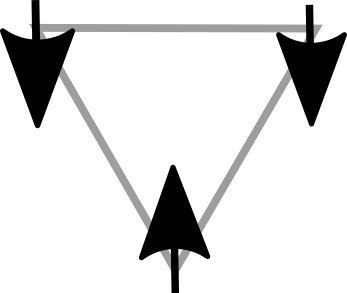}}}) - E(\vcenter{\hbox {\includegraphics[scale = 0.4]{triangles/ppp.png}}}) \bigg) \nonumber \\
    & \hspace{-1.5 cm} = 4 \log 1 = 0.
    \label{eq: mutualInfo_p=0}
\end{align}
The factor of 2 comes from the periodic boundary condition in the spatial direction of the quantum model. The diagrams in the third line are the classical configurations close to the last time slice of the random circuits (FIG.\ref{fig: brickwall}) and the dark regions mark the downward spin clusters. The corresponding quantum state is a product state in spite of the infinite number of on-site degrees of freedom, which is expected when the system only scrambles and dissipates, wherein the unitary scrambling accelerates the dissipation since the local information dissipates globally. 

When $q<\infty$, the mutual information receives correction in $p$, $\Gamma$ and $\frac{1}{q}$, which turns out to be negative. This is not prohibited because R\'{e}nyi entropies do not satisfy subadditivity. Its negativity is heuristically an extreme case of the ``decoupling principle" \cite{Li_Fisher_2021_stat_qec_PRB, Li_Vijay_Fisher_21_arxiv}, where quantities like $\Tr{\rho_{AB}^2}$ may encode more information than $\Tr{\rho_{A}^2}$ or $\Tr{\rho_{B}^2}$ combined, because the former contains interactions between the subsystem $A$ and $B$. This may especially be true for systems that interact with a bath since measuring the subsystem $A$ reveals ``less information about $B$ than 0", where 0 is the decoupling principle in the $\Gamma \to 0$ limit. The naive R\'{e}nyi mutual information can be computed in terms of Boltzmann weights of the individual plaquette (FIG.\ref{fig: Boltzmann weights})
\begin{align}
\tilde{I}^{(2)}_{A:B} (p=0,q,\Gamma>0) = 
   2 \log\left(\frac{w (\uparrow \downarrow \uparrow) w (\downarrow \uparrow \uparrow) }{w (\downarrow \downarrow\uparrow) w (\uparrow\uparrow\uparrow) } \right),
\end{align}

The case $(p=0,\Gamma=0)$ is excluded because in this limit $w(\downarrow \downarrow\uparrow) = w (\uparrow \uparrow \downarrow)=0$, which means the horizontal domain wall costs infinite energy, and the picture breaks down. On the other hand, for $\Gamma = 0$, the mutual information will be in the volume law phase because there will be no domain walls for homogeneous boundaries and the energies of $\vcenter{\hbox {\includegraphics[scale = 0.3]{triangles/mmm.png}}}$ and $\vcenter{\hbox {\includegraphics[scale = 0.3]{triangles/ppp.png}}}$ are 0:
\begin{align}
    & \tilde{I}^{(2)}_{A:B} (p=0,q,\Gamma = 0) \nonumber \\
    & \hspace{-1cm}= F\left( \vcenter{\hbox {\includegraphics[scale = 0.5]{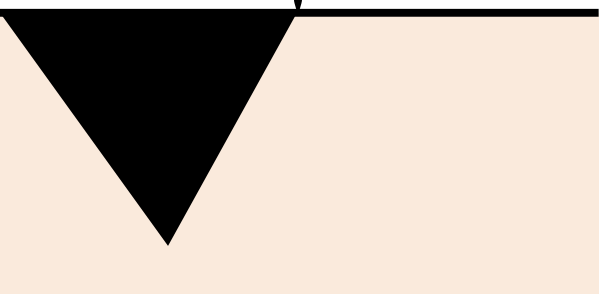}}}\right) + F\left(\vcenter{\hbox {\includegraphics[scale = 0.5]{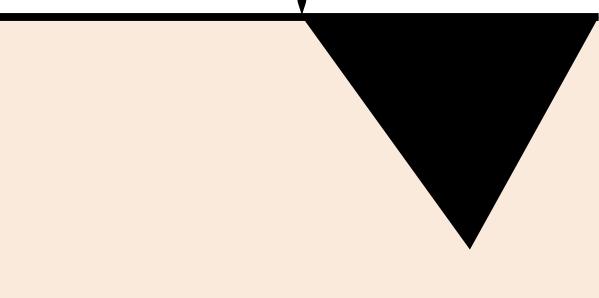}}} \right) \nonumber \\
    & - F\left( \vcenter{\hbox {\includegraphics[scale = 0.5]{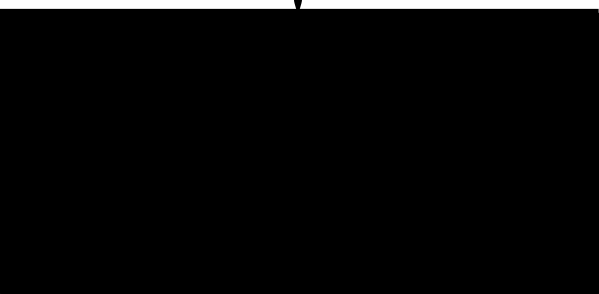}}}\right) - F \left(\vcenter{\hbox {\includegraphics[scale = 0.5 ]{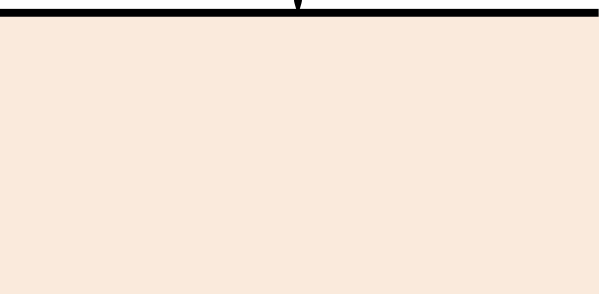}}}\right) \nonumber \\
    & \hspace{-1cm} \approx  - N \log\left(\frac{w (\uparrow \downarrow \uparrow) w (\downarrow \uparrow \uparrow) }{w (\uparrow \uparrow\uparrow) w (\downarrow\downarrow\downarrow) } \right) \nonumber \\
    & \hspace{-1cm} = -2 N \log w(\uparrow \downarrow \uparrow)
\end{align}
where $-2 \log w(\uparrow \downarrow \uparrow) $ would be the prefactor of the volume law growth for the mutual information. 

\subsubsection{\texorpdfstring{$p>0$, $t<t^*_A$}{Lg} (short time regime)}
\label{subsec: short time}

Introducing measurements makes the system more entropic, lowering the free energy of the classical model and the entanglement in the quantum model, but the free energy is still dominated by energy ($p<p_c$). As seen from section~\ref{sec: analytical sigma- channel} and FIG.\ref{fig: Boltzmann weights}, the measurement probability $p$ decreases the strength of the bias field and three-spin interactions.

\begin{figure}
    \centering
    \includegraphics[width = 8 cm]{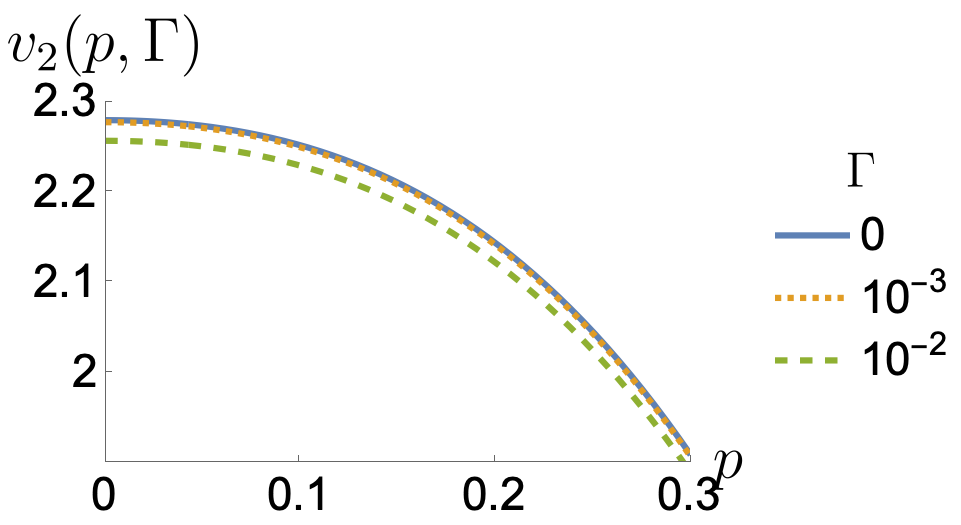}
    \caption{The prefactor of the extensive part of the second R\'{e}nyi mutual information of the dissipative quantum circuit.}
    \label{fig: volume law prefactor}
\end{figure}

We analyze the domain wall configuration  in $F_{A}$ by studying the energetic and entropic contributions of two symmetric random walkers that start from the top ($t=t_{\mathrm{final}}$) $AB$ boundary. First, we eliminate the possibility of a horizontal domain wall because it incurs a high energy cost, as seen from the Boltzmann weights in FIG. \ref{fig: Boltzmann weights}, and will hence be unstable to the following fluctuation
\begin{equation}
    \vcenter{\hbox{\includegraphics[scale = 0.4]{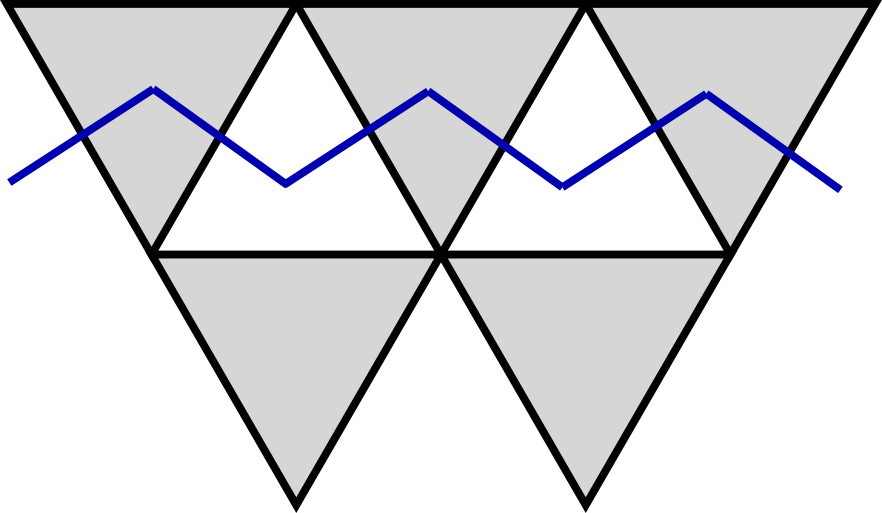}}}
    \Rightarrow\vcenter{\hbox{\includegraphics[scale = 0.4]{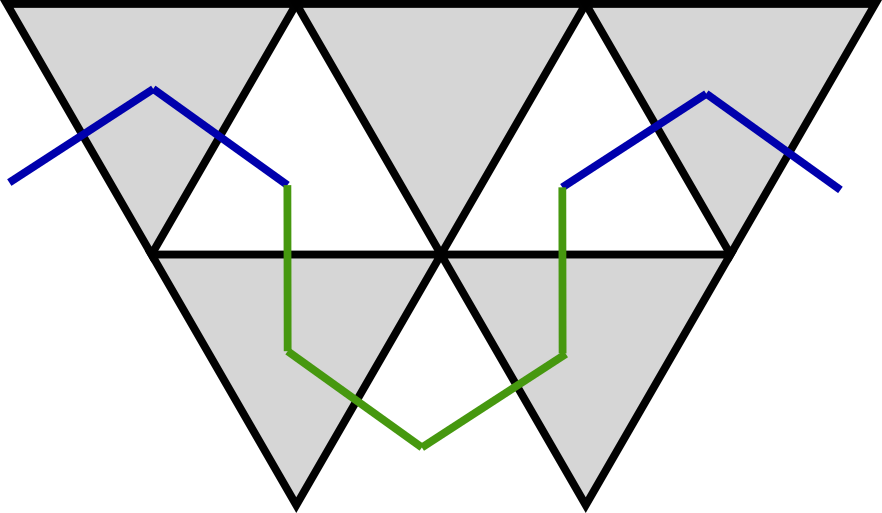}} }
    \label{eq: horizontalDomainFluctuation}
\end{equation}
as the latter configuration is energetically much cheaper. Therefore, the allowed moves for the walkers will be \leftTurn and \rightTurn at short time scales, which means that the domain wall must connect the top $t = t_{\rm max}$ and bottom ($t=0$) boundaries. Within time $t$, the total number of sample paths is $2^{2t}$.  However, not all walks have equal energies, and we estimate the energy of the two parallel walks as it is the most entropic choice amongst the $2^{2t}$ paths while keeping the energy identical, which makes up a macrostate with well-defined free energy. The parallel constraint halves the number of paths, which will underestimate the domain wall fluctuation so the actual entropy of the state is bounded between $2t \log 2$ and $t\log 2$. We therefore denote the entropic correction as $\lambda t$ with $\log(2) \lesssim \lambda \lesssim 2\log(2)$. We note that for much larger $\Gamma$, domain wall configurations must instead minimize the domain area with anti-aligned spins, but we focus here on $\Gamma \ll p < 1$. Similarly, $F_{AB}$ and $F_{\emptyset}$ must both remain fully polarized as a horizontal domain for $F_{AB}$ is too costly, and hence increase linearly with time. Collectively, the mutual information grows linearly, which follows from the free energy difference per unit time $\Delta E$ as:
\begin{align}
     \Delta E(p,\Gamma) =   -\log{\frac{w (\uparrow \downarrow \uparrow) w(\uparrow \downarrow \downarrow)}{w(\uparrow \uparrow \uparrow) w(\downarrow \downarrow \downarrow)}}
\end{align}
This needs to be corrected by the entropy growth $t \log 2$ per random walker. Therefore, the total mutual information growth reads 
\begin{align}
    \tilde{I}^{(2)}_{A:B} &=  F\left( \vcenter{\hbox {\includegraphics[scale = 0.5]{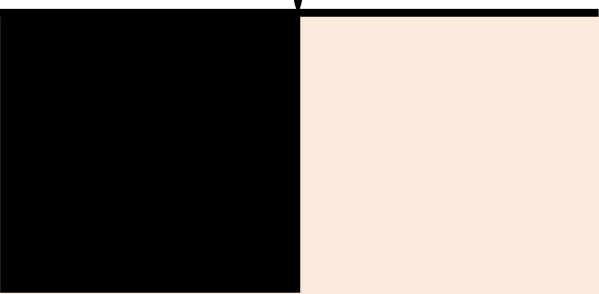}}}\right) + F\left(\vcenter{\hbox {\includegraphics[scale = 0.5]{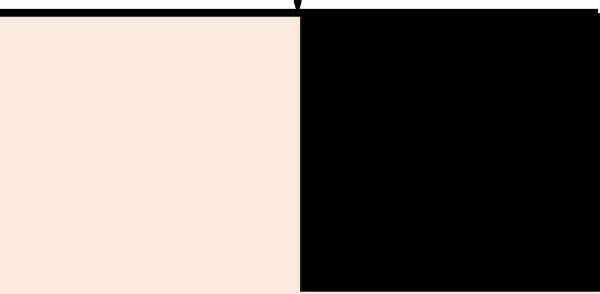}}} \right) \nonumber \\
    & \hspace{.5cm} - F\left( \vcenter{\hbox {\includegraphics[scale = 0.5]{diagrams/F_AB_whole.png}}}\right) - F \left(\vcenter{\hbox {\includegraphics[scale = 0.5 ]{diagrams/F_empty_whole.png}}}\right) \nonumber \\
    & = 2t \Delta E -  2\lambda t \nonumber \\
    &= v_2(p, \Gamma) t .
    \label{eq: velocity}
\end{align}
Termed the entanglement velocity \cite{Nahum_Zhou_2019_PRB}, $v_2(p, \Gamma)$ is shown in FIG.\ref{fig: volume law prefactor} as a function of dissipation and measurement rates for parallel walkers $\lambda = \log(2)$, which shows that they both smoothly slow down the entanglement growth. The growth in free energy here is the same mechanism as the entanglement growth of the R\'{e}nyi entropy in the case without dissipation, but twice in value. Each R\'{e}nyi entropy has its own velocity $v_n$ that is the prefactor of the ballistic surface growth predicted by the KPZ equation, followed by the subleading correction $t^{1/3}$ that accounts for the fluctuation \cite{Li_Vijay_Fisher_21_arxiv, Li_Fisher_2021_stat_qec_PRB, Nahum_Zhou_2019_PRB, Weinstein_2022_PRL,Liu_Li_Zhang_Jian_Yao_arxiv}. A bound on the growth of the mutual information would be difficult to estimate because the speed at which $S^{(n)}_A$ and $S^{(n)}_{AB}$ are separately bounded and with opposite signs. In principle, the estimated velocity for R\'{e}nyi index $n>1$ can be negative at large $p$; however, the above domain wall analysis breaks down for large $p$ as entropy dominates the free energy. 

\subsubsection{\texorpdfstring{$p>0$, $t^*_A <  t < t^*_{AB}$}{Lg} (intermediate time regime) }
\label{subsec: intermediate time}

For $F_A$ and $F_B$ at longer times, the pair of domain walls that traverse the time direction will instead transition to one that traverses the spatial direction, i.e. it starts and ends on the same boundary.
First, the two random walkers can now cross via low-energy diagonal moves and hence terminate. 
Second, the domain with spins aligned in the direction disfavored by the symmetry-breaking terms $h$ and $J_{123}$ will be minimized for longer times, making a domain wall that starts and ends on the top boundary energetically favorable.

\begin{figure}
    \centering
    {\includegraphics[width = 8 cm]{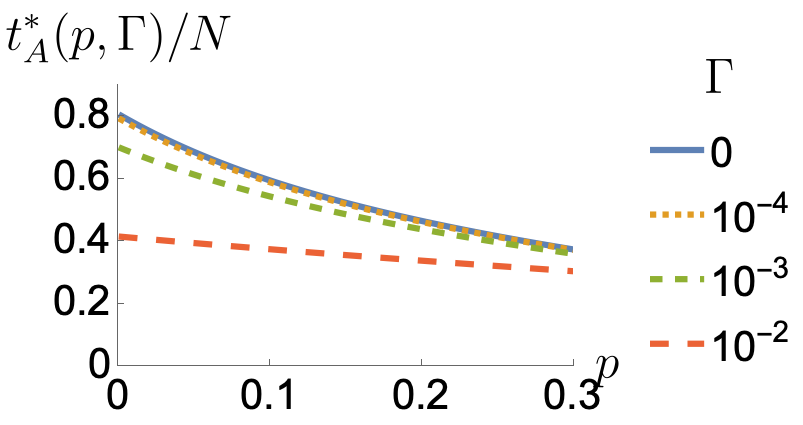}} 
    \caption{The rescaled crossover times $ t^*_A/N $ for finite system size $N$ (right). The system size used is $N=250$.}
    \label{fig: S_A saturation time}
\end{figure}

When the system size $N$ is sufficiently small, the free energy that includes the entropy $\lambda t$ in the previous subsection is appropriate since $t$ for this configuration can be on the same order as $N$. After $t_A^*$, a different domain wall configuration for $F_A$ can appear, which consists of diagonal domain walls that encloses a V-shaped region of $\downarrow$ like that in $F_A(p=0, \Gamma=0)$. Entropic contributions are suppressed as the following fluctuation on the laterals of the triangles
\begin{equation}
    \vcenter{\hbox{\includegraphics[scale = 0.4]{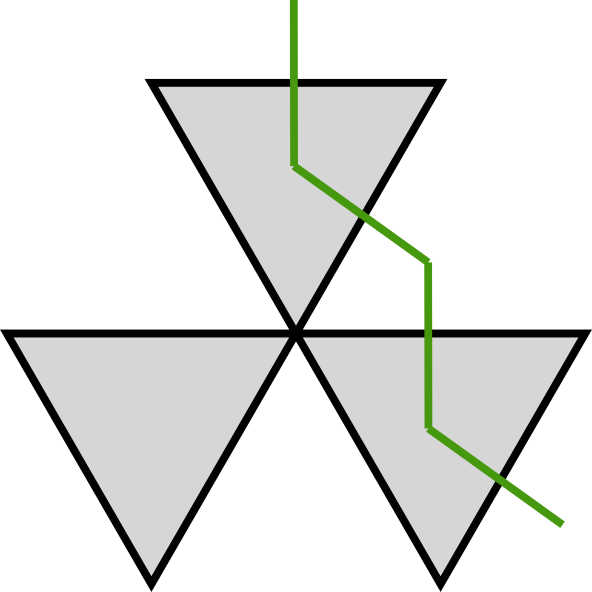}}}
    \Rightarrow\vcenter{\hbox{\includegraphics[scale = 0.4]{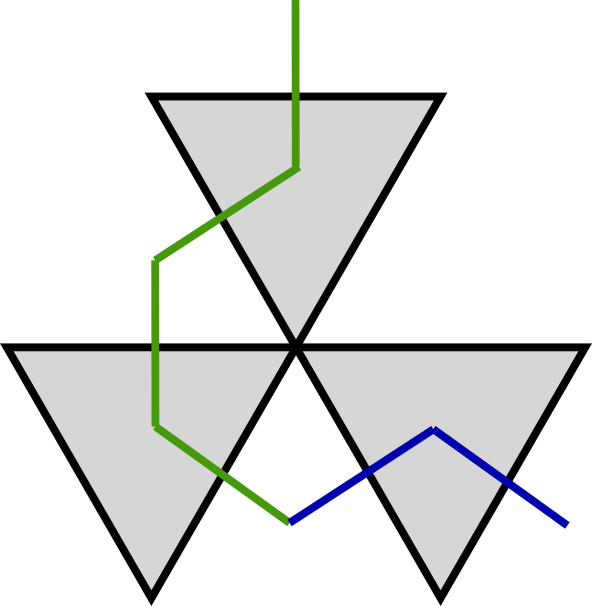}} }
\end{equation}
incurs an enormous energy cost for small $p$ due to substituting \Boltzmann{ppp} by \Boltzmann{mmp}.

Equating the free energies of the short and intermediate-time domain wall states
\begin{align}
    &  \hspace{1cm} F \left(\vcenter{\hbox{\includegraphics[scale = 0.4]{diagrams/F_A_half.png}}} \right) = F \left(  \vcenter{\hbox{\includegraphics[scale = 0.4]{diagrams/F_A_tri.png}}} \right)
\end{align}
defines the time scale $t_A^*$ via the solution of
\begin{align}
    & - t^*_A \bigg( \log w( \uparrow \downarrow \uparrow) w( \uparrow \downarrow \downarrow) + \frac{N-2}{2}  \log w( \downarrow \downarrow \downarrow)+ \lambda \bigg) \nonumber \\& =  - \frac 1 2 \left(\frac{N}{2}-1 \right)^2 \log w(\downarrow \downarrow \downarrow) - N \log w(\downarrow \uparrow \uparrow) \nonumber \\
    & \hspace{1cm} - \bigg[\frac{N}{2} t^*_A - \frac 1 2 \left( \frac{N}{2}+1 \right)^2  \bigg] \log w(\uparrow \uparrow \uparrow)
\end{align}
where only the region where the two saddle points differ is equated in the free energy expression above which consists of $\frac{N t^*_A}{2} $ plaquettes; the rest of the state would be uniformly $\uparrow$. The solution of $t^*_A$ is plotted in FIG.\ref{fig: S_A saturation time}. This $t^*_A$ is proportional to the system size $N$. $t^*_A$ is shortened with increasing $p$ and $\Gamma$. This together with the decrease in the entanglement velocity $v_2$ predicted in \ref{subsec: short time} explains why circuits with higher rate of measurement and dissipation form less entanglement even at short times. We note that in the thermodynamic limit $N \to \infty$ as well as in the large $\Gamma$ limit, or when $N \Gamma \gg 1$, the domain wall will stay closer to the top boundary because the free energy scales with the area of $\downarrow$ which is proportional to $Nt$, and $t^*_A$ will instead be system size independent.

\subsubsection{\texorpdfstring{$p>0$, $t>t^*_{AB}$}{Lg} (long time regime)}

Another dynamical crossover in the mutual information happens when the saddle point of the homogenous $\downarrow$ boundary turns from all $\downarrow$ to having a small island of $\downarrow$ close to the top, enclosed by V-shaped vertical boundaries. This happens in a finite-size system and the time scale can be estimated by setting the free energies of the two configurations to be equal.

For a finite system, the vertical domain walls plus \Boltzmann{mmm} can cost less energy than the horizontal domain wall so one can imagine a typical state is one that zig-zag along the waist, but if there are more zig-zags, there must be more \flatTurn that are costly. We approximate the energetic cost of such a configuration with a single V-shape,  
\begin{align}
    &\hspace{1cm} F\left(\vcenter{\hbox{\includegraphics[scale = 0.4]{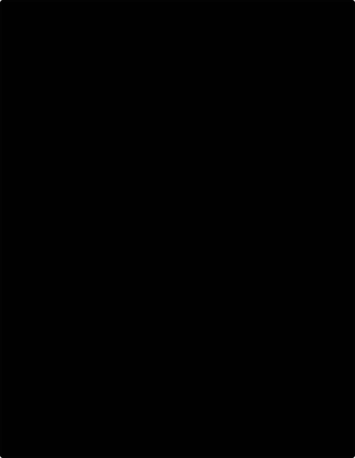}}} \right) = F \left(  \vcenter{\hbox{\includegraphics[scale = 0.4]{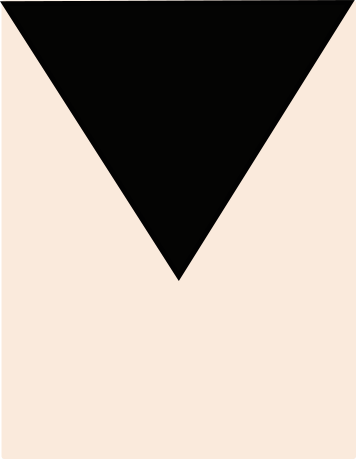}}} \right)
\end{align}
which defines an equation for $t_{AB}^\star$
\begin{align}
    & \hspace{1cm} - N t^*_{AB}  \log w( \downarrow \downarrow \downarrow) \nonumber \\ \nonumber
    &= \frac{(N-2)^2}{2} \log w(\downarrow \downarrow \downarrow) -(2N-4) \log w(\uparrow \downarrow \downarrow) \nonumber \\
    & \hspace{0.5 cm} -2 \log w(\downarrow \downarrow \uparrow) - \left(N t^*_{AB}  - \frac{N^2}{2} \right) \log w(\uparrow \uparrow \uparrow)
\end{align}

\begin{figure}[ht]
    {\includegraphics[width = 8cm]{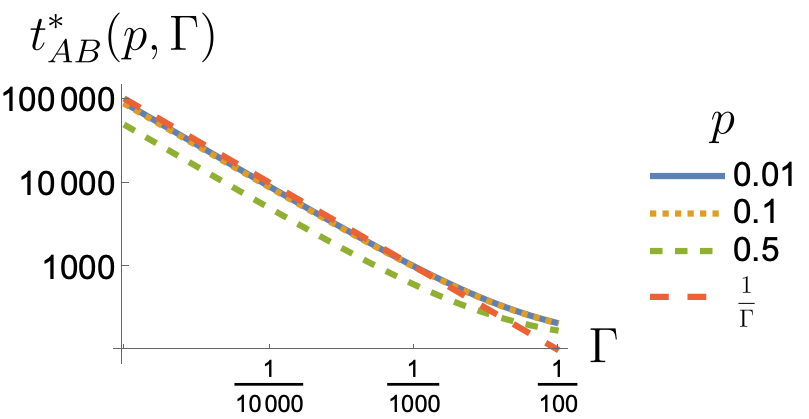} }
    \caption{The crossover time $t^*_{AB}$ at various probability of measurement $\Gamma$, which is independent of the system size. The size used to produce the plot is $N=250$, As $N$ increases, the curve starts to flatten and deviate at large $\Gamma$. The dashed red line ($1/\Gamma$) is a guide to the eye.}
    \label{fig: S_AB saturation time}
\end{figure}

Its dependence on $\Gamma$ and $p$ is shown in FIG.\ref{fig: S_AB saturation time}. The leading terms in $t^*_{AB}$ consist of $\frac{N}{2} + \frac{2 ( \log w (\downarrow \uparrow \uparrow) - \log w(\downarrow \downarrow \downarrow) )}{\log w(\downarrow\downarrow\downarrow) - \log w(\uparrow\uparrow\uparrow) }$ where the second term scales with $1/\Gamma$. For finite systems at a small dissipation rate, i.e. when $1 \gg \Gamma N$, the second term will dominate, and $t^*_{AB}$ scales with $1/\Gamma$ with a weak dependence on $N$. This is the limit when most of the energy cost in $F_{AB}$ comes from the domain wall, which agrees with the analysis in dissipative model without measurement \cite{Li_Sang_Hsieh_2022}. However, when $\Gamma N \gg 1$, the first term will dominate and $t^*_{AB}$ will plateau at $N/2$ when $\Gamma$ surpasses $1/N$. This happens when the energy incurred by the expanding area populated by \Boltzmann{mmm} becomes larger than that of the diagonal domain walls. This is the only allowed saddle point when the dissipation only occurs on the boundary \cite{Li_Sang_Hsieh_2022}; without measurement, the horizontal domain wall is forbidden. We are assuming similarly that for a finite system, the energy cost of a horizontal domain wall outweighs that of a $\downarrow$ region. We will consider the horizontal domain wall in the thermodynamic limit in the next discussion session. Note that $t^*_{AB}$ is not defined for $\Gamma =0$ because in that limit $S_{AB}=0$ at all times.

\begin{figure}[h!]
\subfloat[]
{\includegraphics[width = 5.7cm]{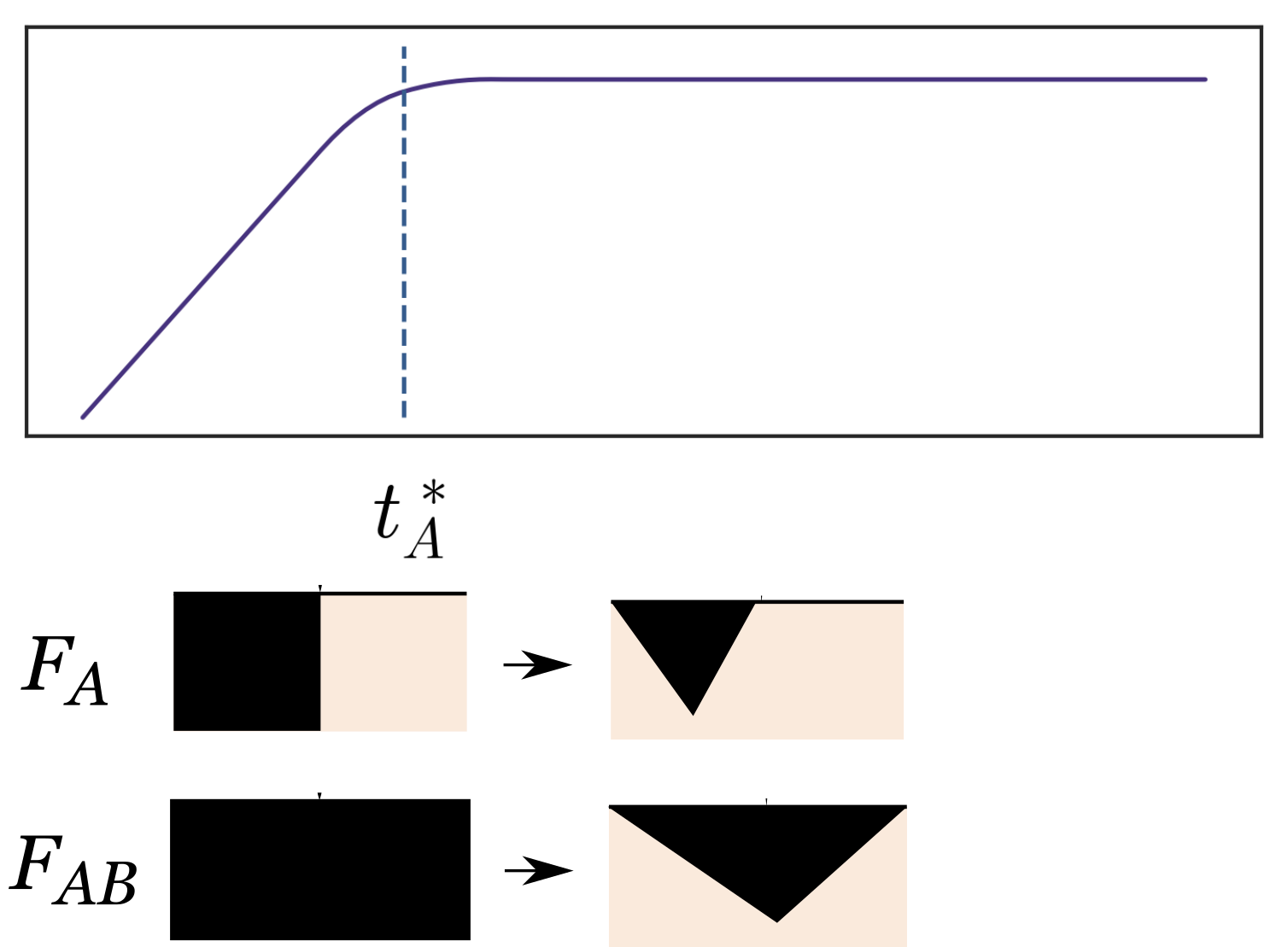}    \label{fig: fN0G}} \\  \vspace{0.2cm}
\subfloat[]
{\includegraphics[width = 5.7cm]{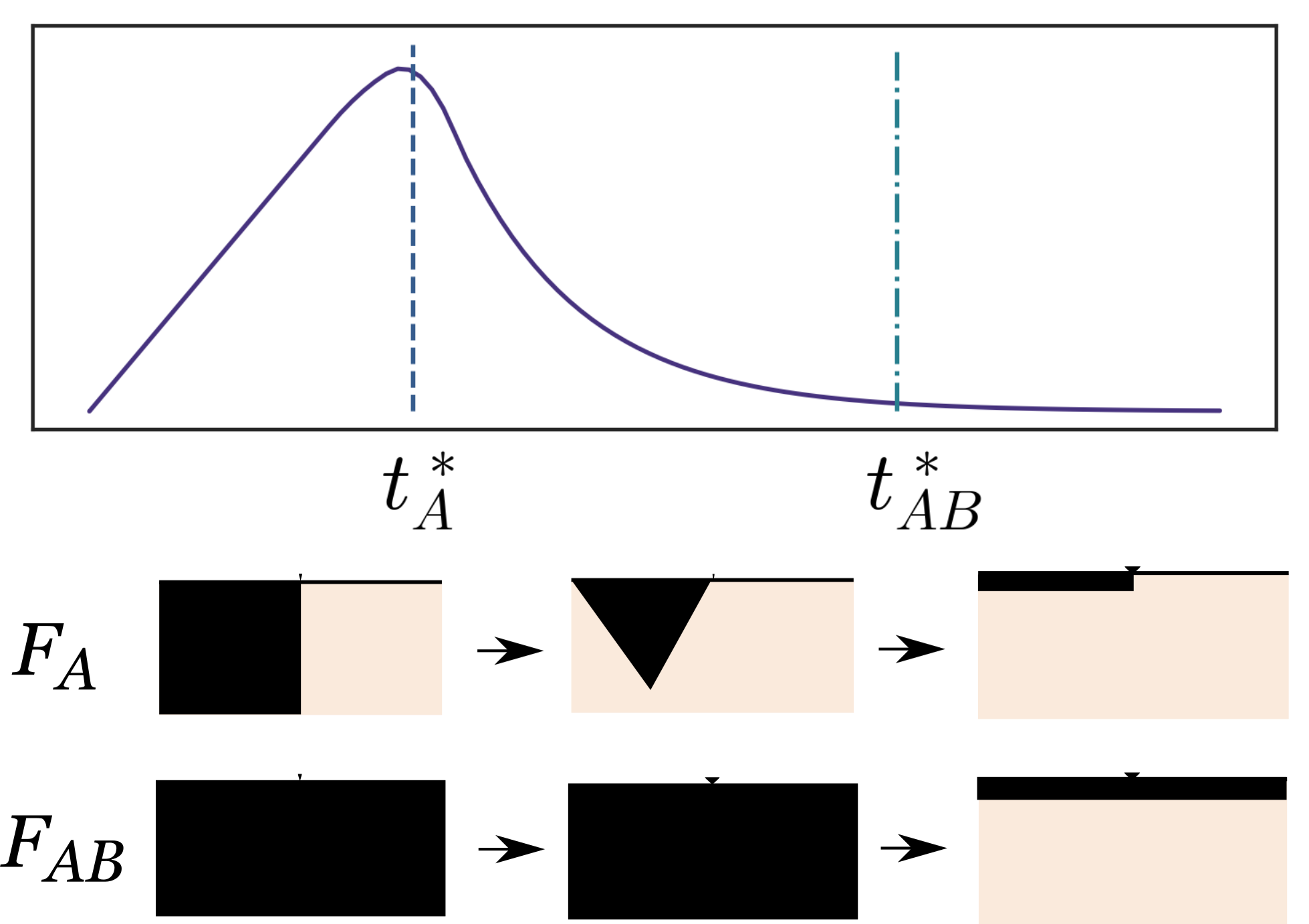}    \label{fig: fN0p}} \\ \vspace{0.2cm}
\subfloat[]
{\includegraphics[width = 5.7cm]{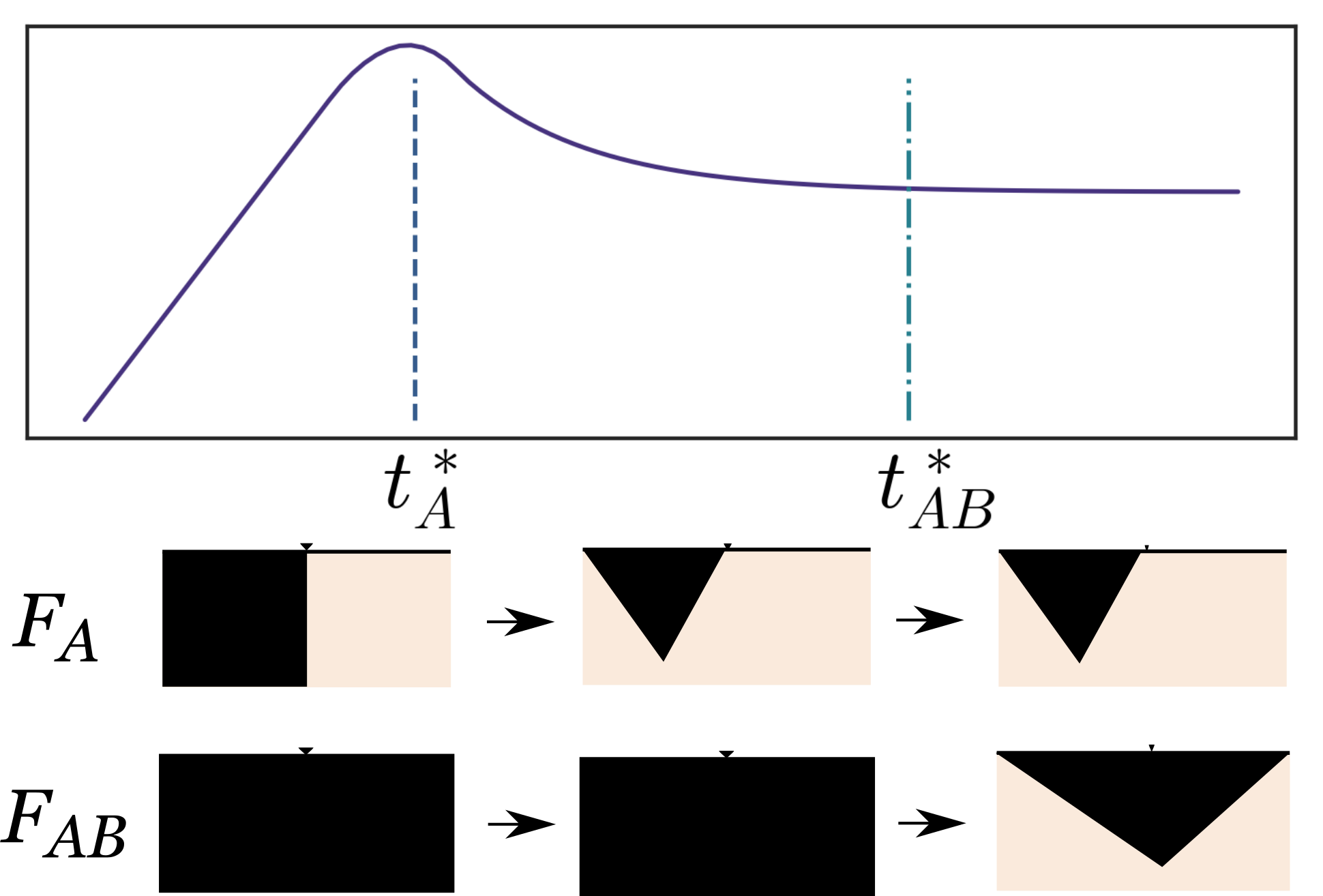}    \label{fig: fNfp}} \\ \vspace{0.2cm}
\caption{The dynamics of the mutual information when (a) $p\geq 0$ and $\Gamma =0$, (b)  $p=0$ and $\Gamma >0$, and (c) $p>0$ and $\Gamma>0$ for finite systems.}
\end{figure}

To summarize the dynamical regimes of finite-size random circuits with or without measurement or dissipation, a key ingredient is the restriction on domain wall growth in both the space and the time dimensions. Measurement and dissipation play very different roles when they are both present in the system; independently however, either will reduce the entanglement of the steady state. When dissipation is absent, one recovers the well-known cases of entanglement growth in (monitored) random unitary circuits, shown in FIG.\ref{fig: fN0G} and entanglement does not decay after a saturation time linear in $N$.

With dissipation but without measurements, entanglement grows ballistically initially as the domain wall stretches in time, but then decays to zero because $F_AB$ grows longer than $F_A$ given that nucleating a domain wall without an appropriate boundary condition is more difficult. In the end, the classical domain wall will stick to the final time slice as it is the most energetically favorable, shown in FIG.\ref{fig: fN0p}. The unitary dynamics does not create more entanglement between the disparate elements within the principal system, and the steady state becomes classical.

With both measurement and dissipation, the initial growth phase does not change. However, measurement and dissipation have competing effects in the statistical mechanical model and on the steady state. The dissipation causes the mutual information to fall off as in the previous case, but the measurement keeps the state from being completely mixed. In the classical model, the measurement leaves some room for $\downarrow$ in a finite system, but compared to the non-dissipative case, its domain wall will be closer to the top boundary due to the biased fields, so the free energy or the mutual information in the quantum model is less than that of the \ref{fig: fN0G} and more than that of \ref{fig: fN0p}, as seen in FIG.~\ref{fig: fNfp}.

\subsubsection{Comments on the time scales as \texorpdfstring{$N \to \infty$}{Lg} }
\label{sec: theory N infinity}
For finite-size systems, we have found two timescales $t^*_A$ and $t^*_{AB}$, where the former depends on $N$. This cannot happen in the thermodynamic limit, because the entanglement should not grow linearly forever when it is coupled to an infinite bath. However, the difference between finite and infinite $N$ is that the parametric energy difference between the domain wall configurations (FIG.\ref{fig: Boltzmann weights}) does not matter at finite times as long as the difference does not increase with the system size. This means that for $F_{AB}$, $\mathcal{O}(1)$ number of diagonal domain walls are allowed as long as it does not grow with the horizontal extent of the system $N$, even if the boundary condition does not induce a domain wall, and the free energy is lowered by the domain walls due to their fluctuations. The energy of the diagonal domain wall which scales with 2$t$ is infinitesimal compared to the bulk uniform \Boltzmann{mmm} which scales roughly with $tN$ and it will be compensated by nucleating a region of \Boltzmann{ppp} around $t=0$ slice that costs less than \Boltzmann{mmm}, which scales with $t^2$. The transition to the steady state will then be the same for $F_{A}$ and $F_{AB}$, both of which will be described by changing from a trapezoid at short times to a hanging horizontal domain wall in the long time limit,
\begin{align*}
    F_A \hspace{0.5cm} \vcenter{\hbox{\includegraphics[scale = 0.5]{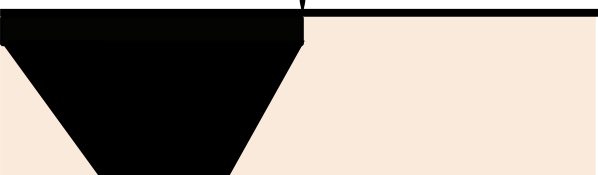}}}
    &\Rightarrow\vcenter{\hbox{\includegraphics[scale = 0.5]{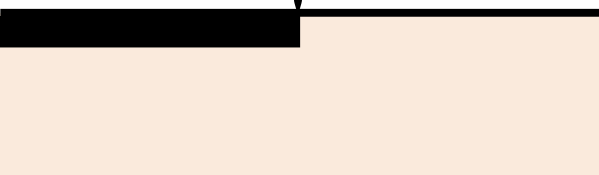}} } \nonumber \\
    F_{AB} \hspace{0.5cm} \vcenter{\hbox{\includegraphics[scale = 0.5]{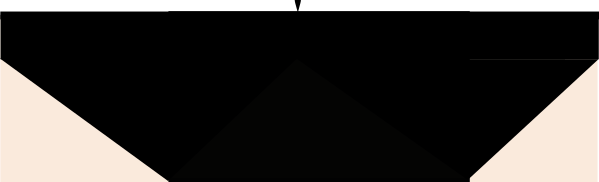}}}
    &\Rightarrow\vcenter{\hbox{\includegraphics[scale = 0.5]{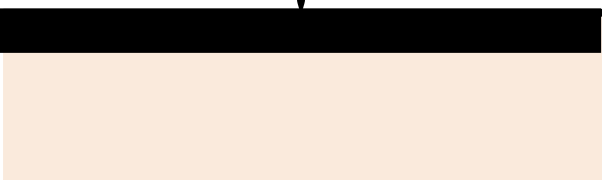}} }
\end{align*}
The time of transition $t^*$ can be obtained by setting the two free energies to be equal and in the large $N
$ limit, it is the same time scale where the $F_A$ in \eqref{eq: velocity} becomes that in \eqref{eq: mutualInfo_p=0} with finite depth and it will be $N$ independent. If we ignore the depth of the hanging horizontal domain wall, we could estimate this by 
\begin{align}
    t^* = \frac{ \log w(\downarrow \downarrow \uparrow) - \log w(\uparrow \uparrow \uparrow)}{\log w(\downarrow\downarrow\downarrow) - \log w(\uparrow \uparrow \uparrow)},
\end{align}
where \Boltzmann{mmm} costs $\Gamma$ more than \Boltzmann{ppp}, so this time scale again scales with $\frac 1 \Gamma$, as expected \cite{Li_Sang_Hsieh_2022}. 

This new configuration also has an impact on the entanglement velocity calculated in \ref{subsec: short time}, which is now time-dependent. The extra diagonal domain walls and the larger \Boltzmann{mmm} area in $F_{AB}$ than that in $F_A$ will reduce the velocity in time and changes the sign of the velocity at a finite, $N$-independent cross-over time $t_s$. This is difficult to estimate as it depends on the number of diagonal domain walls/trapezoids, even if it is constrained to be $\mathcal{O}(1)$, and on the depth at which the trapezoids emerge. A scenario in which the sign of the velocity can change is the following
\begin{align*}
    F_A \hspace{.5cm} \vcenter{\hbox{\includegraphics[scale = 0.5]{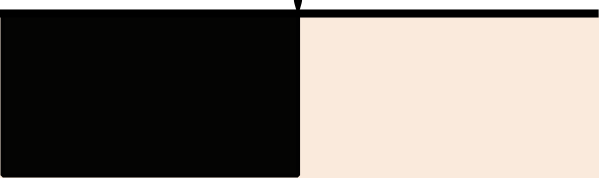}}} \xRightarrow{t_s}
    \vcenter{\hbox{\includegraphics[scale = 0.5]{diagrams/F_A_trapez.png}}}
    &\xRightarrow{t^*}\vcenter{\hbox{\includegraphics[scale = 0.5]{diagrams/F_A_bdry_Ninfty.png}} } \nonumber \\
    F_{AB} \hspace{.5cm} \vcenter{\hbox{\includegraphics[scale = 0.5]{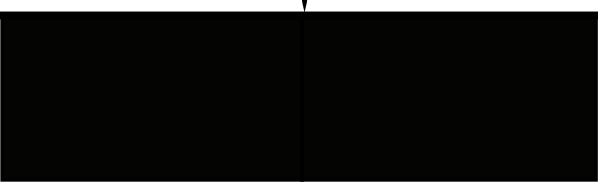}}}  \xRightarrow{t_s}\vcenter{\hbox{\includegraphics[scale = 0.5]{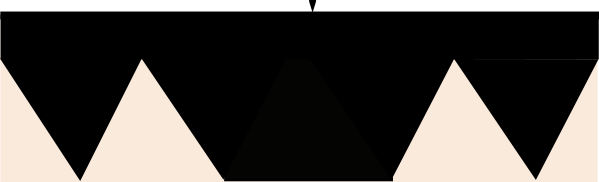}}}
    &\xRightarrow{t^*}\vcenter{\hbox{\includegraphics[scale = 0.5]{diagrams/F_AB_bdry_Ninfty.png}} }
\end{align*}
where the top row describes the state changing in $F_A$ and the bottom row describes the state changing in $F_{AB}$. The left frame occurs at the extremely short time $t<t^*_A$. The middle frame describes a decrease of the mutual information as the number of domain walls in $F_{AB}$ outnumbers that in $F_A$ and $F_{B}$ combined. This change is accompanied by enhanced fluctuations as vertical domain walls will merge areas of the same spin in the vicinity of each other. Nevertheless, this argument about the relation between the two timescales $t_s<t^*$ provides an intuitive picture of how area law entanglement in the steady state is reconciled with ballistic short-time entanglement growth in an infinite system. The entanglement dynamics should be comparable to that in FIG.~\ref{fig: fNfp}, with $t_A$ substituted by $t_s$ and $t_{AB}$ by $t^*$.

While the above rough estimations of crossover times for small systems largely ignored entropic contributions, these are needed for computing precise values of the mutual information in the steady state. Prior work on monitored unitary circuits has mapped the resulting fluctuating domain walls onto a directed polymer problem, with projective measurements acting like random attractive potentials $V(x,t)$ in the bulk \cite{Li_Vijay_Fisher_21_arxiv, Li_Fisher_2021_stat_qec_PRB}. The kinetic energy of these polymers comes from the thermal fluctuation that we have ignored. The boundary condition plays a vital role in computing mutual information. If the mixed boundary condition does not change the bulk state of the lattice magnets, e.g. the completely polarized state, or the nature of the polymers, the mutual information $\tilde{I}^{(2)}_{A:B}$ will only keep an $\mathcal{O}(1)$ area law contribution. In the statistical mechanical description, this means that the polymers in $F_A$ and $F_B$ would be treated as two halves of the polymer in $F_{AB}$, and  $F_A+F_B-F_{AB}-F_{\emptyset} = -\tilde{I}_{A:B} \leq 0$, since entropy increases when a substance is divided into multiple parts. This is not the correct assumption, because the polymer in $F_{AB}$ is a closed contour in the bulk, and cutting it will result in two free floating polymers with much larger entropy than the polymers pinned at the two ends in $F_A$ and $F_B$.  If the energy still roughly cancels up to $\mathcal{O}(\log N)$ between the classical models with different boundary conditions, we quote the free energy difference between the two types of polymer in capillary wave theory that is $\frac 5 2 \log N$ \cite{Li_Fisher_2021_stat_qec_PRB} in the case where measurement is absent. However, the final mutual information does depend on $p$ and $\Gamma$ in \cite{Weinstein_2022_PRL} and our section~\ref{sec: numerics} which will require revisions to explain the steady state entanglement in a monitored dissipative system. We note that in our case, dissipation acts like a long-range gravitational attraction to the top boundary that wants to minimize the area of $\downarrow$ enclosed. This long-range potential controlled by $\Gamma$ will compete with the other localized potential $V(x,t)$ controlled by $p$ in determining the surface tension, and hence the saddle points. A refinement of the classical theory is therefore required to have analytic control over the steady state mutual information.

\section{Numerical simulations}
\label{sec: numerics}
To check the qualitative dynamical trends as a function of $p$ and $\Gamma$ for the quantum system, we now simulate the monitored random-unitary + amplitude damping circuit for small qubit chains up to $N = 14$ following FIG.\ref{fig: brickwall} using  exact many-body density matrix trajectories. At every layer, we first apply the Haar random unitary matrix to the two nearest neighboring qubits, then evolve individual qubits by the Kraus operators for the dissipative channel. Subsequently, each qubit is subjected to a projective measurement at random times, governed by the probability $p$. The circuit can be considered as a Trotter decomposition of a monitored qubit chain that slowly dissipates into a Markovian environment. To improve the numerical efficiency of the density matrix evolution, we optimize tensor contraction between gates and qubits which halves the run time compared to brute force matrix multiplication for the open N-body quantum systems. The dimensionless quantity $\Gamma$, the probability of making measurements $p$ and the system size $N$ determine dynamics and steady-state entanglement, which we study by computing the von Neumann mutual information
\begin{align}
    I_{A:B}(\rho) &= S(\rho_A) + S(\rho_B) - S(\rho)
\end{align}
via simulating the per-measurement-sequence density matrix trajectories and averaging over measurements. $\{A,B\}$ is an equal size partition of the system with periodic boundary condition.
Complementary behavior for the logarithmic negativity is shown in Appendix.\ref{sec: negativity}. Results are presented in the parameter space $(p,\Gamma)$ where $p\in [0,0.6]$ and $\Gamma \in [10^{-4},10^{-2}]$. The range of probabilities encompasses the critical $p_c$ that have been observed and predicted for pure-state trajectories with monitored measurements \cite{Nahum_Chalker_2013_PRB, Pixley_et_al_2020_PRB, Yimu_et_al_2020_PRB}. We focus on weak but non-negligible dissipation rates; for instance, $\Gamma = 10^{-3}$ has been used to test error propagation in Noisy Intermediate Scale Quantum (NISQ) devices \cite{Cirac_2022_PRXQuantum}. While simulable system sizes $N$ remain too small to resolve critical points and scaling exponents for pure-state dynamics, the central object of interest in our work is the competition between dissipation and monitored measurements in governing the short-time entanglement dynamics, to identify peak regimes of mutual information growth, useful for optimizing entanglement generation in quantum systems in a noisy environment. The behavior of the dephasing channel is similar to the amplitude damping channel at small $\Gamma$, which we will not analyze in this work. The difference kicks in only at large $\Gamma$ when the qubit chain completely thermalizes with the environment.

 \begin{figure}
    \centering
    \includegraphics[width=8cm]{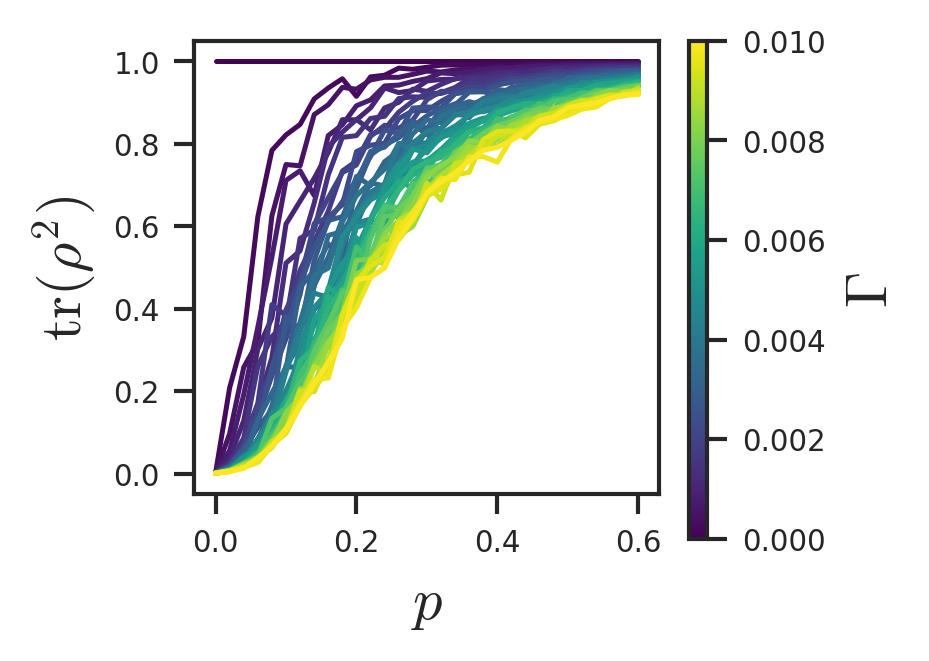}
    \caption{The dependence of the purity of the total steady state on the probability of measurement for different $\Gamma$.}
    \label{fig: purity_vs_p}
\end{figure}

 We choose an initial pure product state, which generically evolves into a mixed state as a function of $\Gamma$ while simultaneously subjected to measurement purification \ref{fig: purity_vs_p}. For large $\Gamma \gg 1$, the state will evolve into the unique ground state of the Lindblad operator. A higher rate of dissipation renders lower purity at a given probability of measurement, as long as $\Gamma<1$, while measurements purify the states (FIG.\ref{fig: purity_vs_p}). Another limit where the steady state is pure is when the dissipation rate is large enough, e.g. $\Gamma>1$, the state will evolve into the unique ground state of $\sigma_-^{\otimes N}$.

\begin{figure}[h!]
    \centering
    \subfloat[]
        {\includegraphics[width = 8cm]{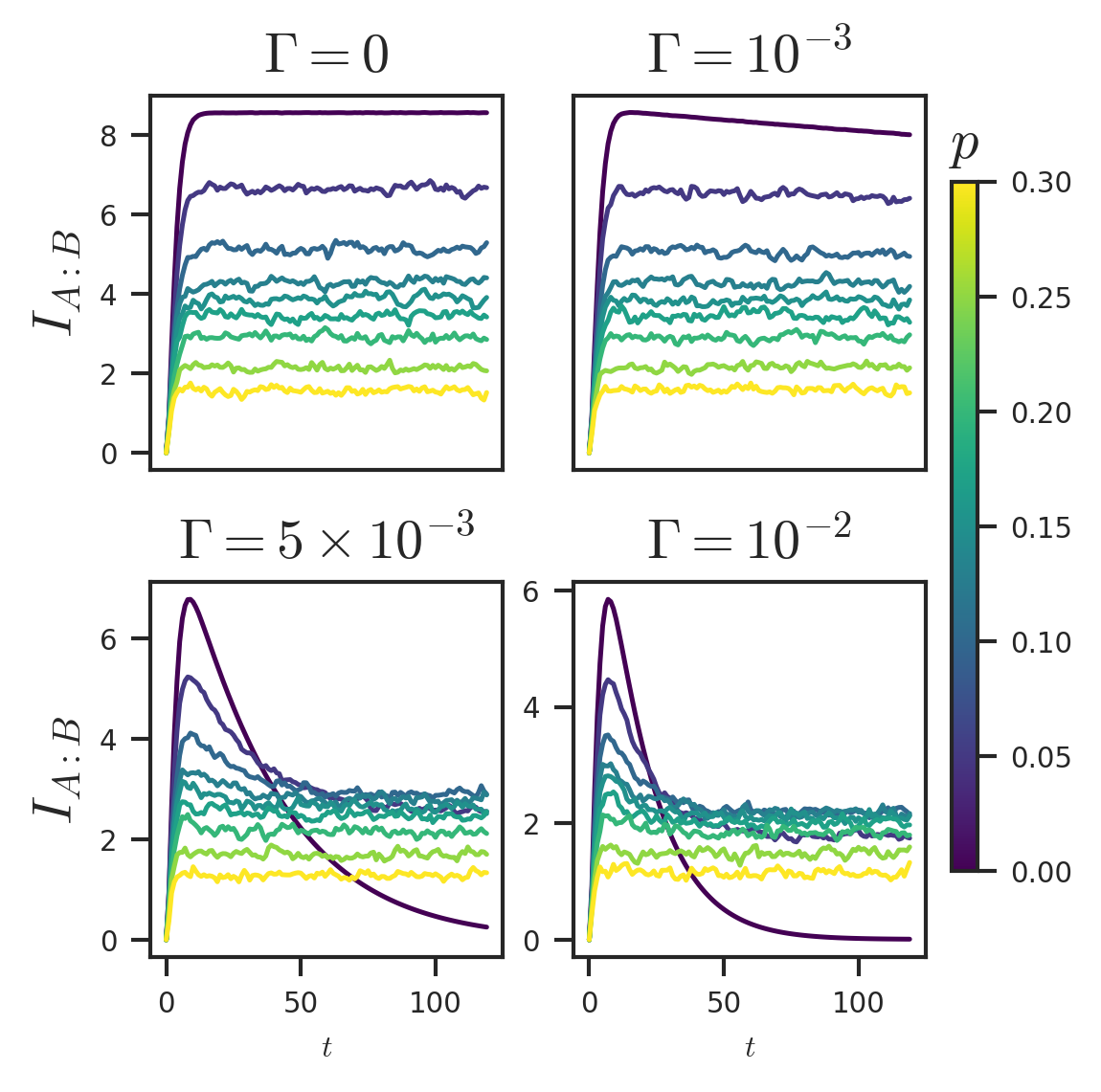}    
        \label{fig: MI_dynamics_N10}}
    \\
    \subfloat[]
        {\includegraphics[width = 8cm]{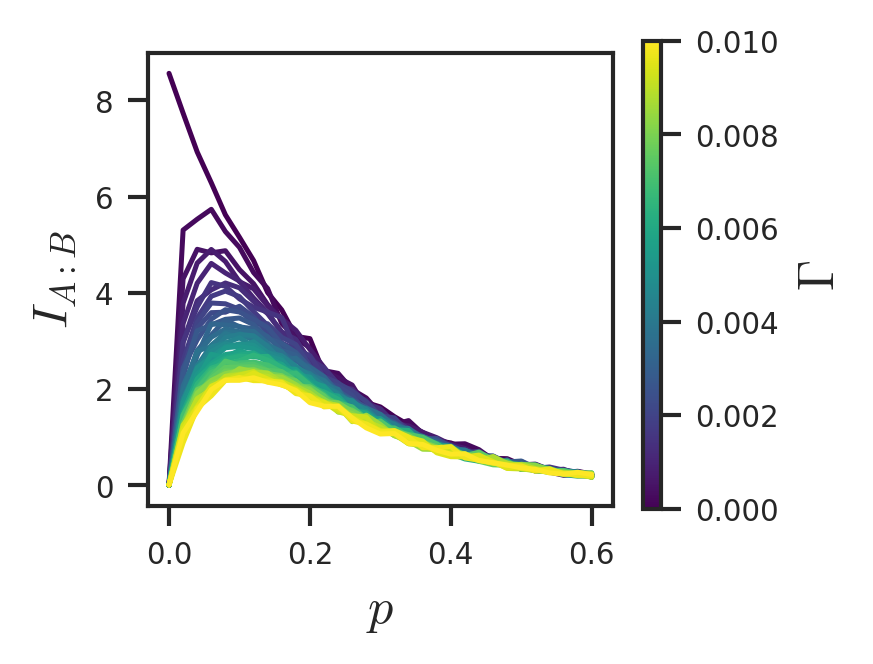} 
        \label{fig: MI_p}}
    \caption{(a) The dynamics of the mutual information within the same time frame at 3 decades of dissipation rates $\Gamma$. The colorbar marks the probability of measurement $p$. The system size is $N=10$ qubits. Each point in time is averaged over 200 independent trajectories over random Haar measure, projective measurements randomly located in space-time, and the random measurement outcomes. (b) The mutual information of the steady state as a function of the probability of measurement with at different values of dissipation rates. The colorbar marks the dissipation rate $\Gamma$.}
\end{figure}

\begin{figure}
    \centering
    \includegraphics[width = 8cm]{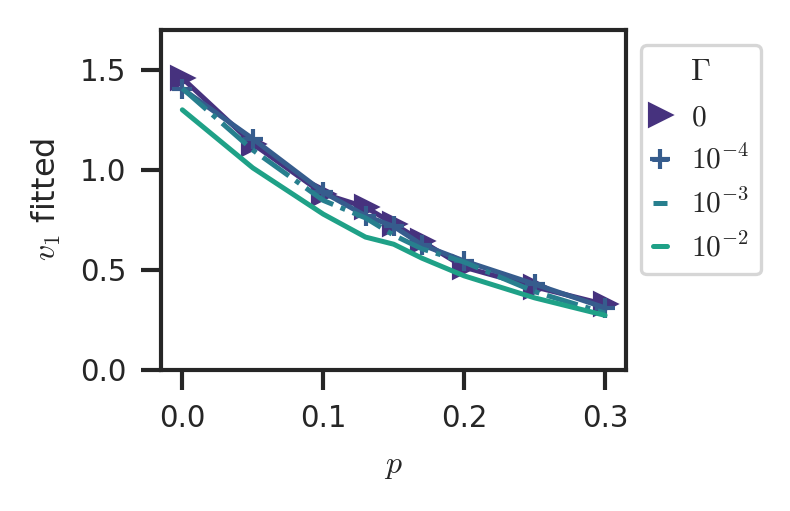}
    \caption{The fitted velocity of increase in the mutual information at short times in system size $N=10$.}
    \label{fig: MI_v}
\end{figure}

The dynamics of the von Neumann mutual information and the negativity agree qualitatively with the picture presented in section~\ref{sec: mincut} from the perspective of the effective classical theory for small systems $N$. At very short times, the entanglement grows ballistically. We fit a linear function to the growth within time $N/2$ (FIG.\ref{fig: MI_dynamics_N10}), estimated by $t_A^*$. The resulting entanglement velocity $v_1$ decreases with increasing $p$ and $\Gamma$, shown in FIG.\ref{fig: MI_v}, similar to $v_2$ predicted in FIG.\ref{fig: volume law prefactor}. The increase in the mutual information terminates smoothly around a time proportional to $t^*_A$ that is shortened by both measurement and dissipation, seen directly from the dynamics (FIG.\ref{fig: MI_dynamics_N10}). Note that the numerical saturation time $t_{\mathrm{sat}}$, estimated in Appendix \ref{sec: negativity}, upper bounds $t_A^*$ in section~\ref{subsec: intermediate time} for fixed $(p,\Gamma)$.

At longer times, the dissipative evolution shows an exponential decay, the time scale of which is controlled by $1/\Gamma$. Larger probability of measurement prevents the steady state from becoming completely mixed while shortening the waiting time. This $1/\Gamma$ waiting time behavior in the Lindblad system again agrees qualitatively with the estimate for the crossover time $t^*_{AB}$ for the second R\'{e}nyi entropy deduced from the classical model, which was shown to be proportional to $1/\Gamma$ and decreases with increasing $p$. This decay describes the thermalization process of the total state $\rho_{AB}$ with the environment after $t_{\mathrm{sat}}$, when most of the entanglement between $A$ and $B$ is formed.   

Finally, in the steady state, the mutual information displays a non-monotonic dependence on the probability of measurement $p$ (FIG. \ref{fig: MI_p}), which decreases the entanglement between $A$ and $B$ however, purifies the final state defined on $AB$. Unlike measurement, dissipation is only destructive in forming long-range correlations between $A$ and $B$ and drives the system towards a mixed state because individual qubits entangle with the environment as opposed to other qubits. The unitary still scrambles the information very quickly within the mixed state but cannot purify the state or retain the quantum correlation from the interaction with the environment. As a result, the mutual information at fixed $p$ monotonically decreases in $\Gamma$. The qualitative dependence of the entanglement as a function of $p$ and $\Gamma$ are consistent with previous findings \cite{Weinstein_2022_PRL}, which includes the peak around $p=0.1$ with varying system size. For the purpose of quantum computing, the relevant time regime with useful entanglement is therefore bounded by $t^*_A$. The estimate of $t^*_A$ for small accessible system sizes is, therefore, the main quantity of interest of this paper.

\begin{figure}[h!]
    \centering
    \hspace{-0.5cm}
    {\includegraphics[width = 8cm]{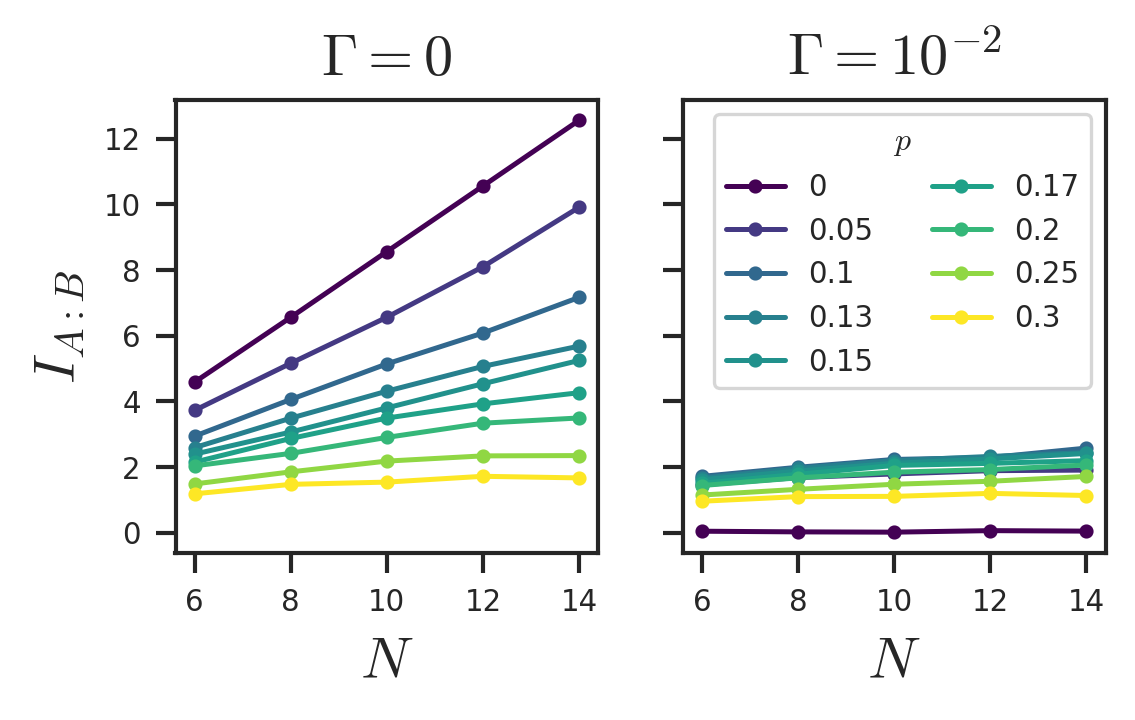}}
    \caption{The scaling of the mutual information with the system size when $\Gamma =0 $ and when $\Gamma =0.01$. } 
    \label{fig: mutualInfo_N}
\end{figure}

Without dissipation, measurement only serves to destroy entanglement, evident from the mutual information curves that descend with increasing probabilities of measurement in the left panel of Fig.~\ref{fig: mutualInfo_N} and the monotonic decreasing negativity in Fig.~\ref{fig: log_neg_p}. However, with dissipation, the steady state lacks any entanglement when no measurement is made, with vanishing mutual information at $p=0$ for all dissipation rates $\Gamma$. As the probability $p$ increases, some coherence of the quantum state is recovered and peaks emerge at the sweet spots where the mutual information is maximized. This entanglement maximizing measurement rate grows slightly with increasing $\Gamma$, and the value of the peak decreases when $\Gamma$ is increased. In the left panel of FIG.\ref{fig: mutualInfo_N}, the dissipationless qubit system is in the volume law phase when $p < p_c \approx 0.16$ and area law when $p > p_c$. In the presence of dissipation, the mutual information obeys an area law in the steady state and stops growing with the system size at large times (FIG.\ref{fig: mutualInfo_N}, right panel). There is no transition in the probability of measurement in the case of random unitary circuits since though the classical analysis where the probability of measurement weakens the symmetry-breaking interactions, but is unable to completely recover the symmetry of the system before the domain walls become deconfined. To the extent that we understand the classical model, there is no symmetry to produce a sharp phase transition.

\section{Conclusion and discussions}
\label{sec: conclusion}
In conclusion, we extend the mapping between random unitary circuits and classical spin models to include both dissipation and measurements. In this mapping, dissipation turns on a bias field and 3-body interactions on downward-pointing triangles in the effective Ising model, and the mutual information can be computed from free energy differences between different boundary conditions. We focus on small and accessible qubit system sizes and qualitatively estimate the short-time, intermediate time and steady-state limits of the entanglement behavior via calculating a generalized second R\'{e}nyi mutual information. While the measurement-induced phase transition is absent at any finite dissipation strength, the Ising effective classical statistical model for two replica is sufficient to qualitatively understand distinct entanglement dynamical regimes. These emerge from distinct domain wall configurations as finite times of the effective Ising model. In classical theory, measurements act to push the model towards a restored Ising symmetry that was broken by dissipation, encouraging the growth of diagonal (vertical) domain walls. Conversely, dissipation serves to push the domain wall towards the pinned boundary, to minimize the area of the spins that anti-align with the biasing terms. The crossover times in the quantum entanglement can be estimated by the change in the saddle point of the classical model as time direction is stretched longer. For finite systems $N$ at low dissipation rate $\Gamma$, the entanglement first grows linearly for a time proportional to $N$ with a $\Gamma$ and measurement probability $p$ dependent velocity, then decays for a time inversely proportional to the dissipation rate $\Gamma$ to reach a steady state with area-law entanglement. 
We finally simulate the open system using exact diagonalization for small system sizes and confirm the qualitative behavior of time scales and entanglement speed that was estimated in the classical theory. We find in agreement with predictions that the short-time dynamics in small systems can create extensive entanglement, which can be usefully targeted for quantum computation, with a bound on achievable time scales estimated from the effective classical model.

An obvious way to improve the presented mapping is to compute quantitative measures of entanglement for mixed states such as negativity and Petz R\'{e}nyi mutual information \cite{Jonah_2023_PRL, Scalet_2021_Renyi}, which satisfies subadditivity and only depends on a single classical state instead of several with different boundary conditions. This requires a general $n$-replicated quantum system to a classical model for arbitrary $n$ and is difficult in the current scheme as the number of spin states increase with the number of replicas, but can in principle, be carried out via a diagrammatic expansion. Another approach is to extend previous analyses using directed polymers by adding long-range potentials that mimic the effect of dissipation. Improved simulations of the open quantum using tensor network\cite{RevModPhys.93.015008} or stabilizer formalism \cite{Li_Sang_Hsieh_2022, Weinstein_2022_PRL} will also permit more refined computations of time scales and transitions.

On a different front, although the tensor network models in holography are still limited in applications so far, the calculation of entanglement entropy is much more tractable in this setup, and it has been fruitful to refine the structure of bulk spatial slices by the entanglement properties of boundary subsystems. The effect of measurement on the boundary BCFT has recently been recognized as information teleportation in the bulk in the context of  MPT \cite{Swingle_et_al_2022_JHEP, Alexey_2022}. Additionally, the recent progress in addressing the black hole information paradox relies on coupling the boundary CFT to a bath in addition to the proposal that AdS/CFT correspondence itself can be understood as quantum error correction has brought brand new insight in to the notion of bulk-boundary mapping \cite{Almheiri_2015, Happy_2015_JHEP}. Our classical analytic picture of the dissipative circuit is the first step towards understanding the bulk theory that is dual to a CFT coupled to a bath.

\begin{acknowledgments}
We thank R. Kamien, J. Kudler-Flam, T. Lubensky, C. de Mulatier, S. Ridout, B. Swingle, Z. Weinstein, and A. Wu for the helpful discussions. CL is supported by the Department of Energy through QuantISED grant DE-SC0020360.
\end{acknowledgments}

\appendix

\section{Kraus operators for generalized dephasing channel}
\label{sec: Kraus der}
The dephasing channel can be written as $$\Phi (\rho) = (1-p) \rho + p \sum_{k=0}^{d-1} \text{tr} (E_{k,k} \rho ) E_{k,k}$$ where $E_{i,j}= \ketbra{i}{j}$ is the matrix unit, $\ketbra{i}{j}$ is in the computational basis and $d=2^N$ is the dimension of the Hilbert space. Its Kraus operators are the eigenvectors of the Choi operator, which can be written as $C_{\Phi} =  \sum_{i,j, i\neq j} \ketbra{ii}{jj} (1-p) + \sum_i \ketbra{ii}{ii} p$. As the column space of the Choi operator is clearly spanned by states of the form $\ket{ii}$, we shorten the repeated index $\ket{ii} \to \ket{i}$. The Kraus operators will be the solution to the equation:
\begin{align}
     (C_{\Phi}(\rho) - \lambda \mathbb{I} ) \ket{\alpha} &= 0 \\
    \Rightarrow  (1-p)\sum_{j} \langle j | \alpha \rangle \sum_i \ket{i} + (p-\lambda) \ket{\alpha} &= 0. \nonumber
\end{align}

The vector equation will be 0 under two circumstances: one is when the vectors $\ket{\alpha} \propto \sum_i \ket{i}$, which is an eigenvector with the eigenvalue $d(1-p)+p$. The other one is when the coefficients of all the vectors are zero, i.e. $\lambda = p$ and $(1-p) \sum_{j}\braket{j}{\alpha} =0$ for any $p$.  Since $\ket{j}$ is summed over an orthonormal basis of the Hilbert space, to make $\sum_j \braket{j}{\alpha} = 0$, $\ket{\alpha}$ needs to be a sum of vectors with coefficients of equal magnitude and opposite signs. Since there can only be $d -1 $ linearly independent eigenvectors in this subspace, without loss of generality, we write this set of eigenvectors with eigenvalue $p$ as $\{ \ket{j}-\ket{0}\}_{j=1}^{d-1}$, to which we must perform the standard Gram-Schmidt orthonormalization procedure.

We use the following strong induction to show that orthonormalizing $\left\{ \ket{\tilde{\lambda}_j} = \ket{j}-\ket{0}\right\}_{j=1}^{d-1}$ gives the generalized diagonal Gell-Mann matrices of $SU(d)$\cite{GellMann}, $$\lambda_ l = \sqrt{\frac{2}{l(l+1)}} \left( \sum_{j=1}^l E_{j,j} - l E_{l+1, l+1} \right),$$ which, in our index convention, are $$\ket{\lambda_l } = \sqrt{\frac{1}{l(l+1)}} \left( \sum_{j=0}^{l-1} \ket{j} - l \ket{l} \right).$$

{\it Proof: } The base case is $\ket{\lambda_1} = \frac{1}{\sqrt{2}} (\ket{0} - \ket{1})$, which satisfies the form.  
Now assume that the form holds $\forall l \leq l'$. Then the $l=(l'+1)$st orthonormal basis vector is:
\begin{align*}
    & \hspace{.5cm} \ket{\dbtilde{\lambda}_{l+1} }  = \ket{\tilde{\lambda}_{l+1} } - \sum_{k=1}^l \braket{\tilde{\lambda}_k }{\tilde{\lambda}_{l+1} } \ket{\tilde{\lambda}_k}\\
    & = \frac{1}{\sqrt{2}} (\ket{l+1} - \ket{0})  \\
    & \hspace{1 cm} +\frac{1}{\sqrt{2}} \sum_{k=1}^l \frac{1}{k (k+1)} \left( \sum_{i=0}^{k-1} \ket{i} - k \ket{k} \right)\\
    & = \frac{1}{\sqrt{2}} (\ket{l+1} - \ket{0})  \\
    & \hspace{0.5cm} + \frac{1}{\sqrt{2}} \left( \sum_{i=0}^{l-1} \ket{i} \sum_{k=i+1}^l \frac{1}{k(k+1)} - \sum_{k=1}^{l-1} \frac{1}{k+1} \ket{k} \right)\\
    & = \frac{1}{\sqrt{2}} (\ket{l+1} - \ket{0})  \\
    & \hspace{0.5cm} + \frac{1}{\sqrt{2}} \sum_{i=0}^{l-1} \left( \frac{1}{i+1} - \frac{1}{l+1} \right) \ket{i} - \sum_{i=1}^{l-1} \frac{1}{i+1} \ket{i}\\
    & = \frac{1}{\sqrt{2}} (\ket{l+1} - \ket{0}) + \frac{1}{\sqrt{2}} \left(\ket{0} - \frac{1}{l+1} \sum_{i=0}^l \ket{i} \right) \\
    &= \frac{1}{\sqrt{2}} \left( \ket{l+1} - \frac{1}{l+1} \sum_{i=0}^l \ket{i} \right).
\end{align*}
Normalizing this by $\sqrt{\braket{\dbtilde{\lambda}_{l+1}}{\dbtilde{\lambda}_{l+1}}} = \sqrt{\frac{l+2}{l+1}}$ gives 
\begin{align*}
    & \ket{\lambda_{l+1}} = \sqrt{\frac{1}{(l+1)(l+2)}} \left( (l+1) \ket{l+1} - \sum_{i=0}^{l} \ket{i} \right).\\
    & \hspace{7cm} \text{q.e.d.}
\end{align*}
Combining this with the other eigenvector with the eigenvalue $d(1-p) + p$, we obtain the expressions of the Kraus operators in section~\ref{sec: setup}.

\section{ \texorpdfstring{Boltzmann weights in the $q \to \infty$}{Lg} limit}
\label{sec: q infinity}

Due to the lack of spin symmetry from dissipation, the number of distinct Boltzmann weights $w_3(\sigma_1,\sigma_2; \sigma_3)$ proliferates rapidly as $n$ is increased. To simplify the analysis, we take the limit of a large local Hilbert space dimension $q\to \infty$, which has previously been used to make the analytical continuation $n\to 1$ tractable \cite{Nahum_Zhou_2019_PRB, Jian_You_Vasseur_Ludwig_2020_PRB, Li_Sang_Hsieh_2022} and recovers the Potts model limit. The Weingarten function has $1/q^2$ suppression in the case when the $\sigma_3$ and $\tau$ on the vertical bonds disagree, so we only keep the contribution from $\tau =\sigma_3$ when computing the three-body Boltzmann weight.  Expanding $w_3(\sigma_1, \sigma_2, \sigma_3)$ in the leading orders of $1/q^2$, $p$, and $\Gamma$ and assuming $1>p \gg \Gamma \gg 1/q^2 >0$, one obtains
\begin{widetext}
    \begin{align}
    \hspace{ 1cm} & w_3^{(2)}(\sigma_1, \sigma_2; \sigma_3) = \sum_{\tau \in \{\uparrow \downarrow\}}  w_g(\sigma_3, \tau) w_d(\sigma_1, \tau) w_d(\sigma_2, \tau) \nonumber \\
    & \hspace{-1.5 cm} = \sum_{\tau \in \{\uparrow \downarrow\}} \left(\frac{1}{(q^4-1)}\delta_{\sigma_3 \tau } - \frac{1}{q^2(q^4-1)} (1-\delta_{\sigma_3 \tau } )\right) \nonumber \\
    & \hspace{-1.3 cm} \times \left(q^2(1-p)^2 \delta_{\sigma_1 \tau} \delta_{\tau \uparrow} + q^2(1-2\Gamma) (1-p)^2 \delta_{\sigma_1 \tau} \delta_{\tau \downarrow} +  q((1-p)^2 + p^2(1-2\Gamma))\delta_{\sigma_1 \uparrow} \delta_{\tau \downarrow}+ q((1-p)^2+p^2)\delta_{\sigma_1 \downarrow}\delta_{\tau \uparrow} \right) \nonumber \\
     & \hspace{-1.3 cm} \times \left(q^2(1-p)^2 \delta_{\sigma_2 \tau} \delta_{\tau \uparrow} + q^2(1-2 \Gamma) (1-p)^2\delta_{\sigma_2 \tau} \delta_{\tau \downarrow} +  q ((1-p)^2+p^2(1-2\Gamma))\delta_{\sigma_2 \uparrow} \delta_{\tau \downarrow}+ q((1-p)^2+p^2)\delta_{\sigma_2 \downarrow}\delta_{\tau \uparrow} \right) \nonumber \\
     & \hspace{-1.5 cm} = \frac{q^4 (1-p)^4}{q^4-1} \left[ \delta_{\sigma_1 \sigma_3}\delta_{\sigma_2 \sigma_3} - 4\Gamma   \delta_{\sigma_1 \sigma_3 }\delta_{\sigma_2 \sigma_3 }\delta_{\sigma_3 \downarrow} \right]  + \mathcal{O}\left( \frac{1}{q} \right) \nonumber \\
     & \hspace{-1.5 cm} = \frac{q^4 (1-p)^4}{q^4-1} \left[ \vcenter{\hbox {\includegraphics[scale = 0.5]{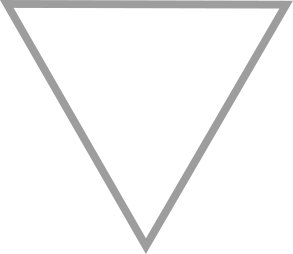}}} - 4\Gamma  \vcenter{\hbox {\includegraphics[scale = 0.5]{triangles/mmm.png}}} \right]  + \mathcal{O}\left( \frac{1}{q} \right),
\end{align}
\end{widetext}
where the empty plaquette means that three spins on the corners are the same. 
Furthermore, only terms $\mathcal{O}(\Gamma)$ are kept.
From the expression above, the Boltzmann weights of \Boltzmann{ppp} and \Boltzmann{mmm} are equal without the $\Gamma$ term, and the highest order replica-symmetry-breaking term makes the weight of $\vcenter{\hbox {\includegraphics[scale = 0.3]{triangles/mmm.png}}}$ smaller by $4\Gamma$. To see how dissipation affects the other spin configurations on a plaquette, one must go to the next order in $1/q$, where the replica symmetric (RS) part is 
\begin{widetext}

    \begin{align}
    & \frac{q^3 (1-p)^2 ((1-p)^2 + p^2) }{q^4-1}( \delta_{\sigma_1 \sigma_3}+ \delta_{\sigma_2 \sigma_3} -2\delta_{\sigma_1 \sigma_3} \delta_{\sigma_2 \sigma_3}) 
\end{align}
and the replica-symmetry-breaking (RSB) part is 
    \begin{align}
        &\frac{q^3 (1-p)^2 }{q^4-1}  \Bigg[ -2 p^2 \Gamma \delta_{\sigma_1 \sigma_3}  \delta_{\sigma_2 \sigma_3} \delta_{\sigma_3 \downarrow}  -2\Gamma \delta_{\sigma_1 \sigma_3} \delta_{\sigma_3 \downarrow}  ( (1-p)^2 + p^2) (1-\delta_{\sigma_2 \sigma_3}) -2 \Gamma p^2\delta_{\sigma_2 \sigma_3} \delta_{\sigma_3 \downarrow})+ (\sigma_1 \leftrightarrow \sigma_2) \bigg] \nonumber \\
        &  = -\frac{2 \Gamma q^3 (1-p)^2 }{q^4-1}  \Bigg[ ((1-p)^2 + 2 p^2) \vcenter{\hbox {\includegraphics[scale = 0.5]{triangles/mmm.png}}} + ((1-p)^2 + p^2) \vcenter{\hbox {\includegraphics[scale = 0.5]{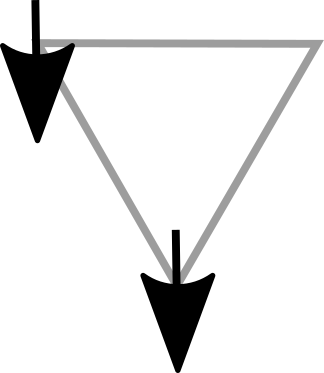}}} + p^2   \vcenter{\hbox {\includegraphics[scale = 0.5]{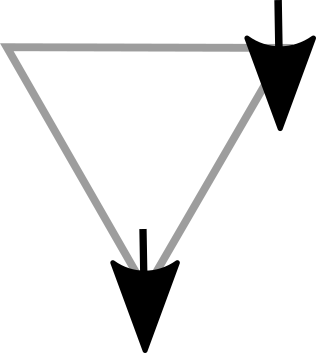}}}  + (\sigma_1 \leftrightarrow \sigma_2) \Bigg],
    \end{align}   
\end{widetext}
where vertices drawn without a spin mean that the spin can point in either up or down, and repeated empty vertices in the plaquette mean that they all take the same spin but can point in either direction. We organize the Boltzmann weights of the other configurations in leading order $\frac{1}{q}$ in table \ref{table: boltzmann}.
\begin{widetext}

\begin{table*}[h!]
\centering 
    \begin{tabular}{ |p{2cm}|p{2cm}|p{2.5cm}|p{3cm}|p{2cm}|} 
        \hline
        \multicolumn{1}{|c}{ domain wall configuration}  &
         \multicolumn{2}{|c|}{ \includegraphics[scale = 0.5]{triangles/domWall0.png}} &
         \multicolumn{2}{c|}{ \includegraphics[scale = 0.5]{triangles/domWall2.png}} \\
          \hline
         spin diagram &
         \multicolumn{1}{c|} {\includegraphics[scale = 0.3]{triangles/ppp.png}}  & 
         \multicolumn{1}{c|} {\includegraphics[scale = 0.3]{triangles/mmm.png}}  & 
         \multicolumn{1}{c|}{\includegraphics[scale = 0.3]{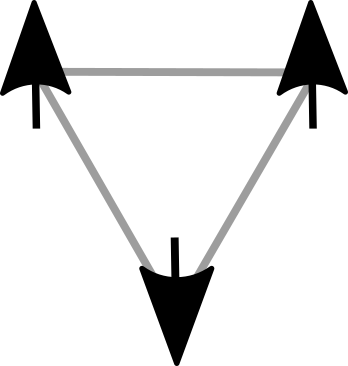}} & 
         \multicolumn{1}{c|}{\includegraphics[scale = 0.3]{triangles/mmp.png}} \\
         \hline
         $w_3^{(2)}$  & $\frac{q^4(1-p)^4}{q^4-1}$ & $\frac{q^4(1-p)^4 (1-4\Gamma )}{q^4-1} $ &  
          {$ \frac{q^2 ((1-p)^2 + p^2  (1-2\Gamma))^2 }{q^4-1} $}& $\frac{q^2 ((1-p)^2 + p^2)^2 }{q^4-1}$\\
          \hline
    \end{tabular}

\begin{tabular}{ |p{2cm}|p{2cm}|p{2cm}|p{2cm}|p{2cm}|} 
    \hline
     \multicolumn{1}{|c|}{ domain wall configuration } & \multicolumn{4}{|c|}{ \includegraphics[scale = 0.5]{triangles/domWall3.png} or  \includegraphics[scale = 0.5]{triangles/domWall1.png} } \\
    \hline
    spin diagram & \includegraphics[scale = 0.3]{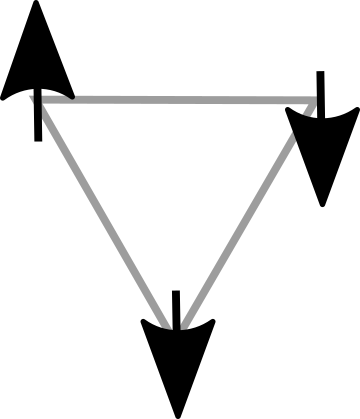} & \includegraphics[scale = 0.3]{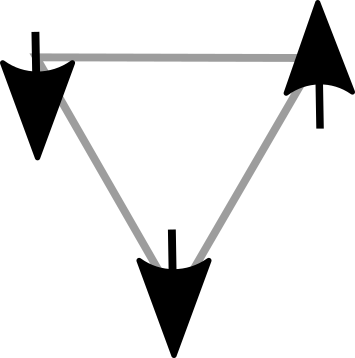} & \includegraphics[scale = 0.3]{triangles/pmp.png} & \includegraphics[scale = 0.3]{triangles/mpp.png}  \\ 
    \hline
    $w_3^{(2)}$  &   \multicolumn{2}{c|}{ $\frac{q^3 (1-p)^2 ( ( 1 -2\Gamma) ((1-p)^2+ p^2) - 2 \Gamma p^2) }{q^4-1} $ }  &  \multicolumn{2}{ c|} {$\frac{q^3 (1-p)^2((1-p)^2 + p^2) }{q^4-1} $} \\
     \hline
     
\end{tabular}
\caption{The Boltzmann weights of domain wall and spin configuration on a plaquette at the leading order of the inverse onsite Hilbert space dimension $1/q$. In the lower table, the reflection symmetry of each triangular plaquette is preserved, but the same domain wall configuration have different weights, depending on the bottom spin orientation.}
\label{table: boltzmann}
\end{table*}

\end{widetext}

Configurations that receive RSB correction at the same order as their RS configuration are the ones with $\tau = \sigma_3 = \downarrow$. Heuristically, this follows from dissipation turning on a magnetic field in the up-spin direction \cite{Weinstein_2022_PRL, Swingle_2021_PRL}. However, we note that the weights account for a combination of a magnetic field pointing in $\downarrow$ direction and a three-body term that favors the $\uparrow$ direction as shown in section~\ref{sec: analytical sigma- channel}. Having no domain wall costs the least energy, which is roughly 0 for \Boltzmann{ppp} or $4\Gamma$ for \Boltzmann{mmm}. The vertical domain wall costs $\log (q)$ if the bottom spin on the plaquette is $\uparrow$, and $\log (q) + 2\Gamma $ if the bottom spin is $\downarrow$. Lastly, the horizontal domain wall costs $2 \log q$ and $2 \log q + 4\Gamma$ with the same dependence on the bottom spin. If the number of $\uparrow$ and $\downarrow$ spins are not equal, then the free energy penalty compared to the homogeneous ground state scales like $N^2 \log (1-4\Gamma) \approx -4 N^2 \Gamma $ where $N^2$ is the size of the 2D lattice.

\section{Scaling of entanglement and negativity of the quantum circuits}
\label{sec: negativity}

As mentioned in the main text, the logarithmic negativity has similar behavior to the mutual information, and it is an actual entanglement monotone, i.e. it does not increase under completely positive trace-preserving maps~\cite{Vidal_Werner_2002}. It is defined as
\begin{equation}
    \mathcal{N}_{A:B} = \log \Tr{\sqrt{||\rho^{T_A}||^2}},
\end{equation}
where $T_A$ means the partial transpose on subsystem $A$. The logarithmic negativity is additive and computes the amount of entanglement distillable. The steady-state logarithmic negativity also forms a peak in probability of measurement $p$, and monotonically decreases as a function of $\Gamma$. An increase in $\Gamma$ from $10^{-3}$ to $10^{-2}$ halves the maximum of log negativity at the $p_{\text{max}} \approx 0.1$. At $\Gamma = 10^{-2}$, the negativity and the mutual information cannot scale linearly with $N$ at any $p$ and the system is deep in the area law phase.

\begin{figure}
    \centering
    \subfloat[]
    {\includegraphics[width =8 cm]{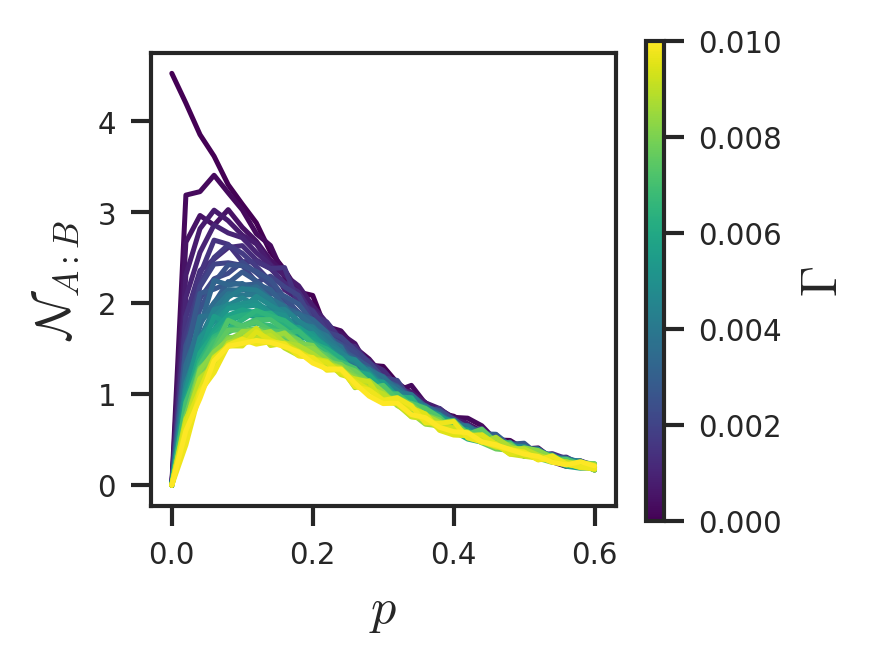}}  \\
    \hspace{-1 cm}\subfloat[]
    {\includegraphics[width =8 cm]{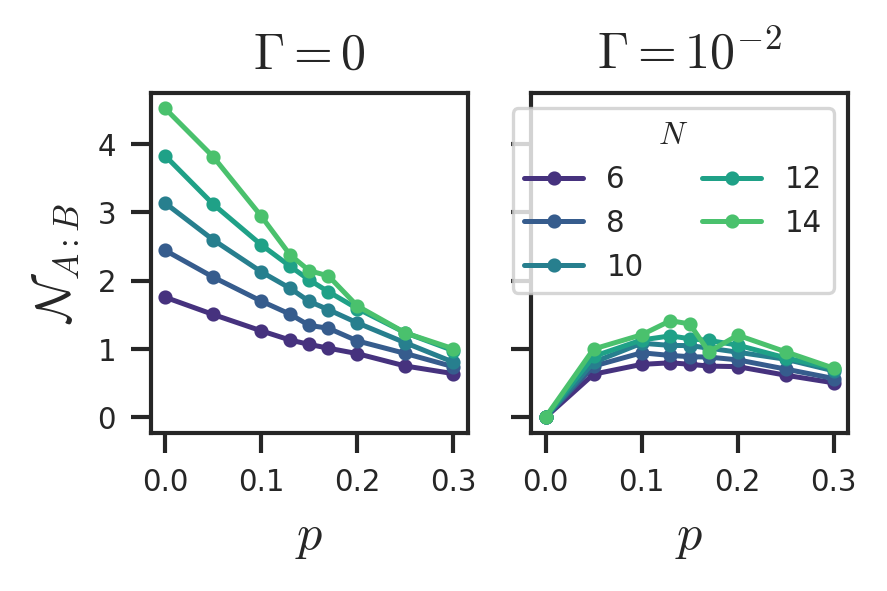}} \\
    \caption{ The dependence of the logarithmic negativity of the subsystem of half the size on the probability of making measurement $p$, where the different colors mark (a) the dissipation rate $\Gamma$ and (b) the different system sizes. In (b), the system size used in the left panel is $N=10$.} 
    \label{fig: log_neg_p}
\end{figure}

Due to the noise in the dynamics and finite system size, we hand pick $t_{\mathrm{sat}}$ where the mutual information saturates or peaks. The transition time is picked such that before $t_{\mathrm{sat}}$, the mutual information monotonically increases in time, and after $t_{\mathrm{sat}}$ the only tolerable increases in mutual information is due to fluctuation. The saturation time has the same trend as $t^*_A$ in section~\ref{subsec: intermediate time}, which decreases with $p$ and $\Gamma$. Due to the small system size, the saturation time is too close to differentiate the $\Gamma=0$ and $10^{-4}$ cases, and as $\Gamma$ increases, the curves of $t_{\mathrm{sat}}$ develop a meaningful difference. We expect that the saturation time will smoothly decay to 0 as a function of $p$ and $\Gamma$.
 
\begin{figure}    
\hspace{-.8cm}
    \centering
    \includegraphics[width = 8 cm]{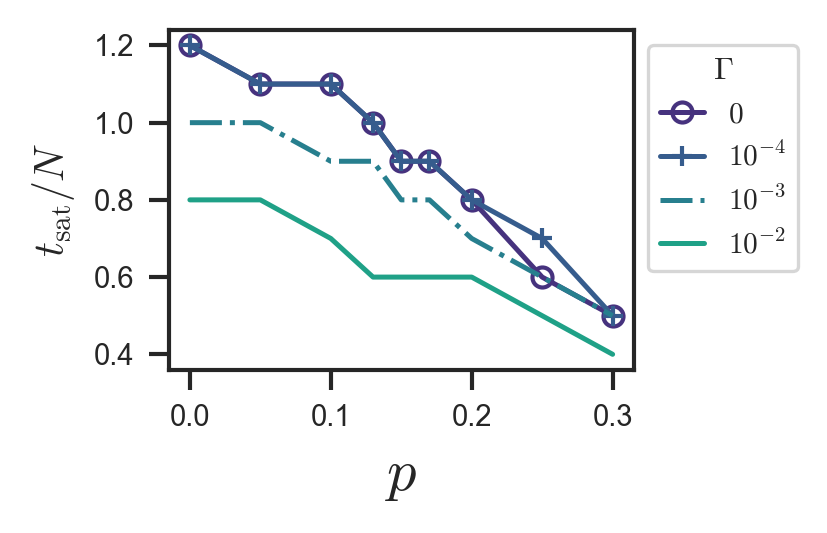}
    \caption{The estimated peak time divided by the system size. The system size giving this plot is $N=10$, so there is not enough time to resolve the saturation time at $\Gamma = 0$ and $\Gamma = 10^{-4}$.}
    \label{fig: MI_sat_time}
\end{figure}

\bibliography{bib}

\end{document}